\documentclass[journal=mamobx,manuscript=article]{achemso}
\setkeys{acs}{articletitle = true}

\usepackage[T1]{fontenc}
\usepackage{amsmath}
\usepackage{graphicx}
\usepackage{mciteplus}

\SectionNumbersOn

\makeatletter

\title{Kremer-Grest models for commodity polymer melts: Linking theory, experiment and simulation at the Kuhn scale}

\author{Ralf Everaers}
\affiliation{
Univ Lyon, ENS de Lyon, Univ Claude Bernard, CNRS, Laboratoire de Physique and Centre Blaise Pascal, F-69342 Lyon, France}

\author{Hossein Ali Karimi-Varzaneh}
\author{Nils Hojdis}
\author{Frank Fleck}
\affiliation{Continental, PO Box 169, D-30001 Hannover, Germany}

\author{Carsten Svaneborg}
\email{science@zqex.dk}
\affiliation{University of Southern Denmark, Campusvej 55, DK-5230 Odense M, Denmark}

\begin{document}
For Table of Content use only:

\begin{center}
{%
\setlength{\fboxrule}{5pt}
\setlength{\fboxsep}{0pt}
\fbox{\includegraphics[angle=270,width=8.3cm]{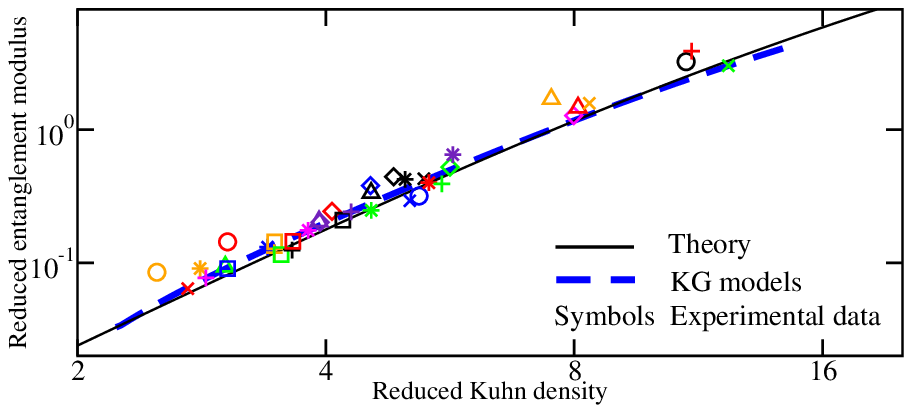}}
}%
\end{center}

\newpage

\maketitle

%150 <= 150 words for Macromolecules

\begin{abstract}
The Kremer-Grest (KG) polymer model is a standard model for studying generic polymer properties in Molecular Dynamics simulations. It owes its popularity to its simplicity and computational efficiency, rather than its ability to represent specific polymers species and conditions. 
Here we show, that by tuning the chain stiffness it is possible to adapt the KG model to model melts of real polymers.
In particular, we provide mapping relations from KG to SI units for a wide range of commodity polymers.
The connection between the experimental and the KG melts is made at the Kuhn scale, i.e. at the crossover from chemistry-specific small scale to the universal large scale behavior.
We expect Kuhn scale-mapped KG models to faithfully represent universal properties dominated by the large scale conformational statistics and dynamics of flexible polymers. In particular, we observe very good agreement between entanglement moduli of our KG models and the experimental moduli of the target polymers. 
\end{abstract}

\maketitle

\section{Introduction}

Polymers are long chain molecules built by covalent linkage of a large numbers of identical monomers~\cite{Flory_53,flory1969statistical}. 
Synthetic polymers are ubiquitous in everyday life due to their unique processing and materials properties.\cite{mccrum1997principles} 
A key problem in polymer science is the relation between structure and dynamics on the molecular scale and the emergent macroscopic material properties. 
The bulk density, the temperature below which the materials become glassy\cite{ngai2007temperature}, or their ability to form semi-crystalline phases\cite{patlazhan2012structural} depend on specific chemical details at the monomer scale. 
Other properties, like the variation of the melt viscosity with the molecular weight of the chains, are controlled by the large scale conformational statistics and dynamics of long entangled chains, which adopt interpenetrating random walk conformations~\cite{DoiEdwards86}. 
Such properties are characteristic of polymeric systems and universal~\cite{deGennes79,DoiEdwards86} in the sense that a large number of chemically different systems exhibit identical behavior, if measurements are reported in suitable, material-specific units. 

The character of the target properties is crucial for making an intelligent choice
of which model to apply in a theoretical or computational investigation. Predicting specific material properties for a
given chemical species often requires atom-scale modeling.
%see e.g. Refs. \cite{theodorou1986atomistic,kotelyanskii1996building,doherty1998polymerization,faller2001local,abrams2003combined,milano2005mapping,sun2005systematic,tsolou2005detailed,tzoumanekas2006atomistic,harmandaris2006hierarchical,neyertz2008molecular}.
A growing body of work aims at developing coarse-grained (CG) polymer
models with lower resolution~\cite{muller2002coarse,baschnagel2000bridging,sun2005systematic,MultiscalePeterKremerSOftMatter2009,li2013challenges}, which are 
designed for specific polymer chemistries such as
polyethylene~\cite{padding2002time,liu2013coarse,salerno2016resolving},
polyisoprene~\cite{faller2002modeling,faller2003properties,li2011primitive,maurel2015prediction},
polystyrene~\cite{harmandaris2006hierarchical,chen2007viscosity,fritz2011multiscale},
polyamide~\cite{karimi2008fast,eslami2011coarse},
polymethacrylate~\cite{chen2008comparison},
polydimethylsiloxane~\cite{maurel2015prediction},
bisphenol-A polycarbonate~\cite{tschop1998simulation,abrams2003combined,hess2006long},
polybutadiene~\cite{strauch2009coarse},
polyvinyl~\cite{milano2005mapping},
and polyisobutylene~\cite{maurel2015prediction}.
Common to these approaches is the selected inclusion of specific chemical
details in the coarse-grained models. They offer insights into which atomistic
details of the chemical structure are relevant for particular non-universal
polymer properties. The inclusion of molecular details is supposed
to preserve a certain degree of transferability, i.e. models optimized to
describe materials at one state point are expected to remain approximately
valid at neighboring state points.~\cite{baschnagel2000bridging,carbone2008transferability,MultiscalePeterKremerSOftMatter2009}
Similarly, careful coarse-graining is supposed to assure representability,
i.e. the ability of a model to predict properties that it was not explicitly
designed to reproduce.~\cite{johnson2007representability}

%
% Dislike Le variable, most people would think in terms of Ne.
%

In contrast, universality is usually taken to justify a ``one-model-fits-all'' approach. 
%to the prediction of universal properties of target materials.
For example, two polymer melts are expected to show identical rheological behaviour, if the (linear) chains have the same {\em effective} length, $Z$ in entanglement units, spatial distances in units of the tube diameter, $d_T$, and time in units of the entanglement time, $\tau_e$. Here $Z=N/N_e$ where
$N$ denotes the chain length and $N_e$ the chain length per entanglement.
This suggests that the linear rheological behavior of a target material can be predicted on the basis of experimental reference data for particularly well-investigated polymer species or using simple, analytically or numerically convenient lattice and off-lattice models, see e.g. refs. \cite{DoiEdwards86,binder1995monte,kremer2000computer,kremer2003computer,kroeger2004simple} for reviews. 
%Universality is often taken to imply that with the appropriate mapping of parameters {\em one} polymer model fits {\em all} melts of intrinsically flexible commodity polymers. 

A prototype example from polymer theory is the Rouse-model of polymer dynamics\cite{rouse1953theory}, which underlies the {\em phenomenological} tube model of entanglement effects in homopolymer melts \cite{DoiEdwards86,likhtman2002quantitative}. 
Simulations targeting generic polymer behavior often employ the bead-spring polymer model introduced by Kremer and Grest (KG)~\cite{grest1986molecular,kremer1990dynamics}, which is also central to the present work. In the KG model approximately hard sphere beads are connected by strong non-linear springs, generating the connectivity and the liquid-like monomer packing characteristic of polymer melts. The interactions are tuned to energetically prevent polymer chains from passing through each other. With the model reproducing the local topological constraints dominating the dynamics of long-chain polymers~\cite{kremer1990dynamics}, non-trivial large scale entanglement properties {\em emerge} through the exact same mechanisms as in real commodity polymer melts.
While the parameters $N_e$, $d_T$ and $\tau_e$ characterizing the entanglement scale are direct input parameters of the tube model, they need to be {\em measured} as a function of the microscopic energy scale, $\epsilon$, bead diameter, $\sigma$,  mass, $m_b$, and time $\tau=\sigma\sqrt{m_{b}/\epsilon}$ of the KG model, before the simulation results can be related to experiment~\cite{kremer1990dynamics}.

Both the standard Rouse/tube and the KG model disregard the polymer contour length as an independent relevant length scale. Theorists assume Gaussian chain statistics, implicitly sending the contour length to infinity in the analytically convenient continuum limit. The contour length of KG chains is finite and well defined, but because Kremer and Grest parameterized their model by mapping it to experimental systems on the entanglement scale, the value is essentially arbitrary. 
To describe situations, where the chains undergo larger deformations, tube models incorporate their contour length as an additional, independent parameter~\cite{DoiEdwards86,likhtman2002quantitative}. 
But how should one control the chain length in KG-like models, where modifications of {\em microscopic} model parameters are bound to influence all emergent mesoscopic time and length scales?

At least qualitatively, finite extensibility was already accounted for in one of the oldest models of polymer physics.
Kuhn's seminal insight in the 1930s was that the large scale conformations of chain molecules can be represented as an  $N_K$ step random walk of ``Kuhn'' segments of length $l_K$ \cite{Kuhn}.
For the proper choice of the Kuhn length, the model reproduces both, the end-to-end distance at full extension, $L=l_K N_K$, and the mean-square end-to-end distance, $\langle R^2 \rangle = N_K l_K^2$, of target polymers. 
While the model obviously needs to be taken with a grain of salt, it provides polymer physics~\cite{deGennes79,DoiEdwards86,RubinsteinColby} with a natural set of microscopic units: the Kuhn length, $l_K$, the Kuhn time, $\tau_K$, as well as $k_BT$ as the natural energy scale in entropy dominated systems. 
Intrinsically flexible polymers exhibit {\em universal} behavior~\cite{deGennes79,DoiEdwards86,RubinsteinColby} beyond the Kuhn scale, while behavior on smaller scales is material specific and dependent on atomic details. For example, the large scale flexibility has completely different microscopic origins in the wormlike chain~\cite{KratkyPorod} and in the rotational-isomeric-state~\cite{flory1969statistical} models. Similarly, there are well-documented exceptions~\cite{Ding_mm_06} to the strong form of the time-temperature superposition principle~\cite{ngai2007temperature}, which postulates identical temperature dependence for microscopic relaxation mechanisms all the way down to the atomic scale. 
Work from the Hassager group\cite{WingstrandAlvarezHuangHassager} underlines the importance of the Kuhn scale for establishing universality in non-linear rheology.
In particular, the authors present {\em three} conditions for {\em non-}linear universality in the rheology of polymer melts \cite{WingstrandAlvarezHuangHassager}. In addition to (i) needing to be composed of chains with the same {\em effective} length, $Z=N_K/N_{eK}$, polymer melts have (ii) to have the same number of Kuhn segments per entanglement, $N_{eK}$, and (iii) to exhibit the same friction reduction in fast elongational flows, in order to show identical behavior.

Here we investigate how the KG model can be used as a convenient tool for exploring universal properties of {\em specific} polymer materials in the above, {\em extended} sense. 
%Can the KG model, which was originally devised to study the generic properties of polymers with an emphasis on computational efficiency, 
%The KG model was not designed to capture the details of any specific chemical polymer species, but rather to describe the generic properties of polymers with an emphasis on computational efficiency. Our objective is to endow KG models with predictive power for universal properties of {\em specific} polymer materials. 
This raises a number of questions:
1) Is there a minimal modification of the standard KG model, that would allow for a coarse-grain description of the full range of standard commodity polymers?
2) How can KG models be related to atomistic simulations, which predict the emerging large scale behavior by accounting for chemical specificity on the atomic scale?
3) How can KG simulation results most easily be compared to theories of polymer physics?
4) How can KG simulations be used to predict the results of experiments performed on real polymers?
5) Is there a price to be paid in terms of computational efficiency in studying material-specific KG models compared to the standard use of the model?

The first question was addressed by Faller and M\"uller-Plathe\cite{faller1999local,faller2000local,faller2001chain}, who introduced a bending potential into the KG model.
Here we show that by tuning the bending potential we can reproduce the full range of effective stiffnesses exhibited by commodity polymers.
In doing so, we rely on results of an accompanying paper, where we have studied the dependence of the characteristic time and length scales in KG bead-spring polymer melts on this parameter\cite{svaneborg2018KGCharacterization}.
%They recognized that real polymers have chemistry specific local interactions
%between monomers that gives rise to a varying degree of stiffness of the polymer
%chains. Hence they they augmented the KG model with an additional potential
%depending on the angle between neighboring bonds, which allows the stiffness
%of KG model polymers to be varied.
%
Our working hypothesis is that questions 2)-4) can largely be reduced to a choice of units or the matching of key characteristic length and time scales. Interestingly, the resulting ``Kuhn scale-mapped KG models'' turn out to be as or even more computationally efficient as the original model.

%
%The purpose of the present paper is to introduce ``Kuhn scale-mapped KG models'' as a convenient tool for exploring emergent universal properties of {\em specific} polymer materials in the above, {\em extended} sense. 
%To obtain valid coarse-grain descriptions for commodity polymer melts, we tune the strength of a bending potential introduced into the KG model by Faller and M\"uller-Plathe\cite{faller1999local,faller2000local,faller2001chain}. 
%The proposed mapping relies on results of a preceding paper, where we have studied the dependence of the characteristic time and length scales in KG bead-spring polymer melts on this parameter~ \cite{svaneborg2018KGCharacterization}.
The paper is structured as follows.
We review the necessary theoretical background in Sec.~\ref{sec:Background}.
In Section~\ref{sec:KuhnScale} we introduce with the Kuhn scale the units of length and time which are central to our scheme for locally mapping real polymers onto Kremer-Grest chains. 
The results cited in the remainder of Sec.~\ref{sec:Background} serve to illustrate, that two monodisperse polymer melts of chemically different polymers are expected to show the same universal large scale properties, provided (i) they are characterized by the same number of Kuhn segments per chain, $N_{K}$, (ii) their densities correspond to the same dimensionless Kuhn number, $n_{K}$,  and (iii) properties are measured in the ``natural'' Kuhn units. 
The actual mapping is discussed in Sec. \ref{sec:matching}. 
In Sec.~\ref{sec:commodity} we transcribe well known results from Fetters {\it et al.}~\cite{fetters94} for the chain structure of a wide range of commodity polymer melts to the Kuhn scale.
Sec.~\ref{sec:KGparameterization} summerizes the results of the accompanying paper, where we have studied the dependence of the characteristic time and length scales in KG bead-spring polymer melts on  the strength of the bending potential~\cite{svaneborg2018KGCharacterization}.
In Sec.~\ref{sec:mappingstatic} we derive mapping relations for static properties.
In particular, we provide tables specifying a one-parameter KG force field for a wide range of experimental polymer melts, i.e. we list 1) which bending stiffness to use for modelling a particular chemical polymer species, and 2) how to translate simulation results expressed in KG units into predictions for the specific polymer material expressed in SI units.
Sec.~\ref{sec:temperature dependence static mapping} discusses the transferability of the force field to other temperatures.
Sec.~\ref{sec:mappingtime} focuses on time-temperature superposition as a means of estimating of how simulation time in our KG models is related to real time.
As a first test, we compare in Sec.~\ref{sec:plateau moduli} plateau moduli inferred from KG models to experimental values.
The discussion in Sec.~\ref{sec:Discussion} focuses on the place of Kuhn scale matched KG models within the multiscale hierarchy of polymer polymers. We propose to view Kuhn scale matching as a special case of structure based coarse-graining and discuss the large effective time step of our models together with the expected speedup relative to atomistic simulations.
Finally, we briefly conclude in Sec. \ref{sec:Conclusion}.

\section{Background}\label{sec:Background}
In this section, we provide a brief outline of polymer theory~\cite{DoiEdwards86,deGennes79,RubinsteinColby,KhokhlovGrosberg}. The point of the exercise is to show that Kuhn scale-mapped KG models can be expected to have predictive power for emergent polymer properties. 

\subsection{The Kuhn scale}\label{sec:KuhnScale}
The Kuhn length, 
\begin{equation}\label{eq:lK}
l_{K}\stackrel{L\gg l_K}{=}\frac{\langle R^{2}\rangle}{L},
\end{equation}
characterizes the crossover from local rigid rod to random walk behavior. 
It is not straightforward to infer the Kuhn length from the chemical structure of a polymer in its melt state as it depends on intramolecular interactions, chemistry-specific local packing, and universal long-range correlations\cite{flory1969statistical,FloryNobelPrize,Wittmer_Meyer_PRL04}.

A known Kuhn length can be used to characterize the large scale structure of polymer melts via two related dimensionless numbers.
The number of Kuhn segments per chain,
\begin{equation}
N_{K}=\frac{\langle R^{2}\rangle}{l_{K}^{2}}=\frac{L^{2}}{\langle R^{2}\rangle},
\end{equation}
is a dimensionless measure of chain length.
If  $\rho_K$ denotes the number density of Kuhn segments, then the number of Kuhn segments within the volume of a Kuhn length cube, 
\begin{equation}\label{eq:nkdef}
n_{K}=\rho_{K}l_{K}^{3},
\end{equation}
provides dimensionless measure of density for polymeric materials. We refer to $n_K$ as  ``Kuhn number''. 
In Kuhn units, the chain density is given by
\begin{equation}\label{eq:rho_chain}
\rho_{c}=\frac{\rho_{K}}{N_K}\ .
\end{equation}

To characterise the dynamics, one can define the friction coefficient, $\zeta_K$, of a Kuhn segment undergoing Brownian motion.  Interpreting $\zeta_K$ as a viscous Stokes drag, $\zeta_K \propto \eta_K l_K$, it is convenient to define an effective viscosity at the Kuhn scale as
%{\bf Please check. Maybe I'm off by a factor of two or something and it would be better to define this with a prefactor of 1/36 to preserve the simple form of eq. (\ref{eq:etarouse})}
%Checked. and it should be 36. sum[p^-2,1,NK] -> pi^2/6 for NK->oo
\begin{equation}\label{eq:etaK}
\eta_K = \frac{1}{36} \frac{\zeta_K}{l_K}\ .
\end{equation}

The fundamental time scale of the dynamics of intrinsically flexible polymers is set by the time that it takes a Kuhn segment to diffuse ($D_{K}=k_{B}T/\zeta_{K}$) over a distance comparable to its own size. Again it turns out to be practical to incorporate some numerical prefactors into the definition of the Kuhn time:
\begin{equation}\label{eq:tauk}
\tau_{K}=\frac{1}{3\pi^{2}}\frac{\zeta_{K}l_{K}^{2}}{k_{B}T} = \frac{12}{\pi^2}\frac{\eta_K l_K^3}{k_BT}.
\end{equation}

The universal aspects of the mesoscale conformations\cite{wittmer2007intramolecular,glaser2014collective}
and liquid structure\cite{clark2013effective,mccarty2014analytical} {\em beyond} the Kuhn scale can often be described by Gaussian chain models. 
% Wittmer: Intramolecular long-range correlations in polymer melts: The segmental size distribution and its moments
% glaser: Collective and single-chain correlations in disordered melts of symmetric diblock copolymers: quantitative comparison of simulations and theory
% Muller: Computer simulation of asymmetric polymer mixtures
% Clark: Effective potentials for representing polymers in melts as chains of interacting soft particles
%{\bf Not sure what's in the papers we cite and if these citations still makes sense!}
% removed the Muller reference, since it is BFM simulations and no relation to liquid structure.
%Added mccarthy An analytical coarse-graining method which preserves the free energy, structural correlations, and thermodynamic state of polymer melts from the atomistic to the mesoscale
This ansatz preserves the information on the mean square chain extensions, $\langle R^{2}\rangle$, without retaining the chain contour length, $L$, as relevant variable. This is particularly apparent in the continuum limit, which is frequently employed in theoretical calculations\cite{DoiEdwards86,Freed}. 

\subsection{Rouse dynamics}\label{sec:RouseDynamics}
We begin our short {\it tour d'horizon} with the Rouse model~\cite{rouse1953theory,DoiEdwards86}, which describes the dynamics of short unentangled polymers. Rouse considered the Langevin dynamics of a ``Gaussian'' chain composed of beads, which experience local friction and which are connected by harmonic springs representing the entropic elasticity of polymer sections beyond the Kuhn scale. In this model, the maximal internal relaxation time of a chain is given by the Rouse time
\begin{equation}\label{eq:taurouse}
\frac{\tau_{R}}{\tau_K}=N_{K}^{2}\ ,
\end{equation}
while the macroscopic melt viscosity can be written as
\begin{equation}\label{eq:etarouse}
\frac{\eta}{\eta_K}
%= \left(\frac 1{18} \frac{\zeta_K}{l_K} \right) n_K \, N_K \ 
= n_K N_K\ .
\end{equation}

In particular, Eqs.~(\ref{eq:taurouse}) and (\ref{eq:etarouse}) illustrate the utility of the natural Kuhn units in expressing emergent universal properties.

\subsection{Packing and the invariant degree of polymerization}\label{sec:PackingInvariantN}
The key for understanding the properties of polymer melts is the realisation, that chains strongly interpenetrate. 
%There are various ways to characterize the mutual chain interpenetration in polymer melts. 
The Flory number, 
\begin{equation}
n_{F}=\rho_{c}\langle R^{2}\rangle^{3/2}=n_K N_{K}^{1/2} \ ,
\end{equation}
is defined as the number of chains populating, on average, the volume spanned by one chain. That the Flory number is large, explains why chains behave nearly ideally in dense melts\cite{flory49} and why such polymer systems can often be well described by mean-field theories~\cite{schmid1998self,muller2005incorporating}. 

The invariant degree of polymerization,
\begin{equation}\label{eq:invariant degree of polymerization}
\bar{N} = \rho_c^2 \langle R^{2}\rangle^{3} =n_K^2 N_{K} = n_{F}^2 \ ,
\end{equation}
is dimensionless measure of chain length {\em and} interpenetration and related to the number of intermolecular pair-interactions a given reference chain experiences.
It plays a key role in more complex polymer systems such as block-copolymers undergoing micro phase separation~\cite{fredrickson87,glaser2014collective}. 
The corresponding packing length~\cite{WittenPacking,fetters1999packing} is defined as
 \begin{equation}\label{eq:packing}
p\equiv\sqrt{ \frac{\langle R^{2}\rangle}{\bar{N}} } = \frac{l_K}{n_K} = \frac1{\rho_c \langle R^{2}\rangle}\ ,
%=\frac{V_{c}}{l_K\langle R^{2}\rangle}=\left(l_{K}^3\rho_{K}\right)^{-1}=n_{K}^{-1},
\end{equation}
such that monomers found at a spatial distance smaller than $p$ from a reference monomer typically belong to the same chain. Monomers found at a larger distance have an increasing probability to belong to a different chain.
%
%It can be interpreted as the spatial distance from a monomer, up to which most other monomers belong to the same chain.
%
%To define a length scale characterizing the chain packing in a polymeric
%material, one can consider a spherical region centered on a chosen monomer.
%If the region is small, then most monomers found inside will belong to the
%same chain as the chosen monomer. If the region is large, then most monomers
%inside the region belong to other chains.
%The packing length~\cite{WittenPacking,fetters1999packing} is related to the Kuhn number as
%\begin{equation}\label{eq:packing}
%\frac{p}{l_K}=\frac{V_{c}}{l_K\langle R^{2}\rangle}=\left(l_{K}^3\rho_{K}\right)^{-1}=n_{K}^{-1},
%\end{equation}
It is evident from Eq.~(\ref{eq:invariant degree of polymerization}) that Kuhn matching is {\em sufficient} to reproduce {\em all} emergent properties, which depend of the invariant degree of polymerization, $\bar{N}$, but by no means {\em necessary}. 
For a more detailed discussion of these aspects in the context of multi-scale modeling, we refer the reader to Refs.~\cite{muller2013speeding,zhang2015communication}.

%For our present purposes, it is important to note that $n_{F}$ and hence the degree of this interpenetration can again be simply expressed as a sole function of the two dimensionless numbers characterising a melt on the Kuhn scale:

%
%For a given polymer material, the Flory number increases with
%the square root of the chain length, since  $\rho_c\propto N_K^{-1}$ and $R\propto N_K^{1/2}$.

%{\bf packing length suppressed for the time being}
%To define a length scale characterizing the chain packing in a polymeric
%material, one can consider a spherical region centered on a chosen monomer.
%If the region is small, then most monomers found inside will belong to the
%same chain as the chosen monomer. If the region is large, then most monomers
%inside the region belong to other chains. The packing length~\cite{WittenPacking,fetters1999packing}, 
%%
%\begin{equation}\label{eq:packing}
%p=\frac{V_{c}}{\langle R^{2}\rangle}=\left(l_{K}^{2}\rho_{K}\right)^{-1}=\frac{l_{K}}{n_{K}},
%\end{equation}
%%
%defines the crossover between these two regimes. It corresponds to the root-mean
%square end-to-end distance of polymers with Flory number $n_{F}\equiv1$ and $N_{K}=1/n_{K}^{2}$.
%For melts of synthetic polymers, the packing length is comparable to the
%monomer size (Tab. \ref{tab:Derived-polymer-parameters})

\subsection{The entanglement scale}\label{sec:EntanglementScale}
%In the following, we take a closer look at dynamic properties.
Chains undergoing Brownian motion can slide past each other, however, their backbones cannot cross\cite{DoiEdwards86}. As a consequence, the motion of long chains is subject to long-lived topological constraints\cite{Edwards_procphyssoc_67}. 
The constraints become relevant at scales beyond the entanglement (contour) length\cite{fetters1999chain,fetters1999packing}, $L_{e}$, or the equivalent number of Kuhn units between entanglements, $N_{eK}=L_e/l_K$. In the present context $N_{eK}>1$.
The corresponding spatial scale is the so called tube diameter,
\begin{equation}\label{eq:dT}
\frac{d_{T}}{l_{K}}=\sqrt{\frac{\langle R^{2}(N_{eK})\rangle}{l_{K}^{2}}}=\sqrt{N_{eK}}\ .
%=\frac{\alpha}{n_{K}}.
\end{equation}

Just like the fundamental time scale $\tau_K$ is determined by the Kuhn segment diffusion coefficient
and the Kuhn length, the diffusion of an entanglement segment ($D_e=k_BT/[N_{eK}\zeta_K]$) and the tube diameter
also define a characteristic time scale $\tau_e$:
\begin{equation}\label{eq:taue_over_tauK}
\frac{\tau_{e}}{\tau_K}=N_{eK}^{2}
%=\left(\frac{\alpha}{n_{K}}\right)^{4}.
\end{equation}

According to the packing argument for loosely entangled polymers~\cite{lin87,kavassalis87}, the ratio of the tube diameter and the packing length is given by a universal constant, $\alpha$. Experimental results~\cite{fetters94} (Tab. \ref{tab:Kuhn}), simulation data~\cite{svaneborg2018KGCharacterization} and a geometric argument~\cite{rosa2014ring} for the local pairwise~\cite{everaers2012topological} entanglement of Gaussian chains suggest
\begin{equation}\label{eq:alpha}
\alpha = \frac{d_T}{p} = 18\pm2\ .
\end{equation}

%A geometric argument~\cite{rosa2014ring}  suggests $\alpha=20.49$ for the local pairwise~\cite{everaers2012topological} entanglement of Gaussian chains. The analysis of large experimental data sets for polymers at $T=413K$ and $T=298K$ yielded $\alpha=19.36$ and $\alpha=17.68$ respectively~\cite{fetters94}. From our Kuhn mapping of the KG model, we obtained $\alpha=17.7$ in the limit of the most flexible chains ($\kappa=-1\epsilon$, $n_K=2.1$).~\cite{svaneborgContiII} {\bf Not clear. Are you referring to a single system? And why do we need Kuhn mapping for this?} For the polymers in Tab. \ref{tab:Derived-polymer-parameters} we obtain an experimental estimate $\alpha=\langle d_{T}/p\rangle=18.0\pm1.3$ consistent with the previous values.
The packing argument implies that there are $\alpha =\frac{\rho_K}{N_{eK}} d_T^3$ entanglement strands per entanglement volume and that the entanglement length is given by 
\begin{equation}\label{eq:netheory}
N_{eK}=\left(\frac{\alpha}{n_{K}}\right)^{2} \ .
\end{equation}

Uchida {\it et al.}~\cite{uchida2008viscoelasticity} developed a scaling theory to describe the crossover to the tightly entangled regime, suggesting instead
\begin{equation}\label{eq:nekuchida}
N_{eK}=x^\frac{2}{5}\left( 1+ x^\frac{2}{5} +x^2 \right)^{\frac{4}{5}}
\quad\mbox{with}\quad x=\frac{\alpha}{n_K}\ .
\end{equation}

For small Kuhn numbers, Eq.~(\ref{eq:nekuchida}) agrees with the packing prediction, but corrections become noticeable for $n_K>10$.

Above the entanglement time, the Rouse model fails to describe dynamic correlations in polymer melts. The universal behavior of entangled chains depends on chain length only through the number
\begin{equation}\label{eq:Z}
Z=\frac{N_{K}}{N_{eK}}\ ,
\end{equation}
of entanglements per chain and is best discussed in the  ``entanglement units'' $d_T$ and $\tau_e$ of spatial distance and time.
A key point to note is that the simple relation $Z=\bar{N}/\alpha^2$ suggested by the packing argument breaks down, when the Uchida corrections become relevant. In this case, the number of entanglements per chain, $Z(n_K)$, and the invariant index of polymerization, $\bar{N}(n_K)$, become {\em different} universal functions of the Kuhn number, $n_K$.

\subsection{The tube model}\label{sec:Tube model}

Modern theories of polymer dynamics~\cite{DoiEdwards86} are based on the idea that entangled chains are confined  to a one-dimensional, diffusive motion (reptation~\cite{degennes71}) in tube-like regions in space~\cite{Edwards_tube_procphyssoc_67}. 
In the limit of long chains the maximal relaxation time~\cite{degennes71} is given by
\begin{equation}\label{eq:taureptation}
\tau_{max}=3Z^{3} \tau_{e}\ ,
%\frac{\tau_{max}}{\tau_K}=3Z^{3}\frac{\tau_{e}}{\tau_K} 
%= 3 \frac{n_{K}^2 N_K^3}{\alpha^2}\ .
\end{equation}
%For chains of finite length, relaxation processes such as contour length fluctuations and constraint release compete with reptation~\cite{DoiEdwards86,likhtman2002quantitative}.
%
%
%Turning now to macroscopic material properties, in slowing down the chain equilibration after a deformation, entanglements dominate the viscoelastic behavior of high molecular weight polymeric liquids. 
and for $\tau_{e}<t<\tau_{max}$ the shear relaxation modulus, $G(t)$ exhibits a rubber-elastic plateau, $G_{N}=\frac{4}{5}G_{e}$, where the entanglement modulus,
\begin{equation}\label{eq:reducedmodulus}
G_{e}=\frac{\rho_{K}}{N_{eK}}k_{B}T \ ,
%\frac{G_{e}l_{K}^{3}}{k_{B}T}=\frac{\rho_{K}l_{K}^{3}}{N_{eK}}=\frac{n_{K}^{3}}{\alpha^{2}}.
\end{equation}
is given by the product of entanglement density and thermal energy.
From the time dependent shear relaxation modulus one can obtain the shear compliance and the
melt viscosity. The asymptotically expected result~\cite{DoiEdwards86} for long entangled chains is given by
\begin{equation}
\eta=\frac{\pi^2}{15} G_{e}\tau_{max}\ .
%\frac{\eta}{\eta_K}=\frac{\pi^2}{15} \frac{G_{e}\tau_{max}}{\eta_K}=\frac{12}5 \frac{n_K^5 N_{K}^{3}}{\alpha^4}\ .
\end{equation}
%%
%suggesting that this property should also be quantitatively reproduced by Kuhn scale-mapped KG models. 

\section{Matching at the Kuhn scale}\label{sec:matching}

In Sec.~\ref{sec:Background} we have identified three relevant length scales: the packing length, $p$, the Kuhn length, $l_K$, and the tube diameter, $d_T$. 
Polymer theory being typically directly formulated in terms of these scales, the results are straightforward to adjust to an experimental target system, for which $p$, $l_K$ and $d_T$ are known.  
But in setting up computational models, we have to deal with the difficulty, that the relevant polymer scales {\em emerge} from interactions and are only indirectly controlled through the parameters of the employed model. Does this mean, that we need to embark on a complicated search for parameter combinations, which allow us to fulfill two independently controlled ratios like $p/l_K$ and $d_T/l_K$? Or maybe one (the standard KG?) model ``fits-all'' and it sufficies to map its predictions to the various experimental target systems?

\subsection{The case for Kuhn scale matching}\label{sec:WhyMatchAtKuhnScale}

As illustrated in Sec.~\ref{sec:Background}, universal static, dynamic, mesoscopic and macroscopic properties of polymer melts emerge from the Kuhn scale. They depend on just two dimensionless parameters: the Kuhn number, $n_K$, characterizing the contour density of the target material and the number of Kuhn segments per chain, $N_K$, as a measure of chain length(s). 
The first parameter is specific for the particular polymer chemistry, while the second characterizes the (polydisperse)
composition of particular melts under investigation.
Specifically, the matching of $n_K$ assures that the model properly reproduces 
(i) the ratio of packing and Kuhn length, $p/l_K$, Eq.~(\ref{eq:packing}),
(ii) the number of Kuhn segments per entanglement length, $N_{eK}$, Eqs.~(\ref{eq:netheory}) and (\ref{eq:nekuchida}),
(iii) the ratio of tube diameter and Kuhn length, $d_T/l_K$, Eq.~(\ref{eq:dT}), and
(iv) the ratio of the Kuhn and entanglement times, $\tau_e/\tau_K$, Eq.~(\ref{eq:taue_over_tauK}).
Matching $N_K$ assures
(i) comparable ratios of average and maximal chain elongation, $\langle R^2 \rangle/L^2 = 1/N_K$, Eq.~(\ref{eq:lK}),
(ii) comparable invariant degrees of polymerization, $\bar{N}$, Eq.~(\ref{eq:invariant degree of polymerization}),
(iii) comparable numbers of entanglements per chain, , $Z$, Eq.~(\ref{eq:Z}),
(iv) comparable ratios of maximal relaxation and Kuhn time, $\tau_{max}/\tau_K$, Eqs.(\ref{eq:taurouse}) and (\ref{eq:taureptation}).
Kuhn scale-mapped KG models can be expected to have predictive power for emergent polymer properties, if we may take it for granted, that this universality of properties of different chemical species also extends to computational models that exhibit the key features of polymer melts: chain connectivity, local liquid-like monomer packing, and the impossibility of chain backbones to dynamically cross through each other.
To most convenient way to make predictions is (i) to express the simulation results for a suitable KG model in Kuhn units of the simulation model and (ii) to convert them to SI-units using the Kuhn length, $l_K$, Kuhn time, $\tau_K$, Kuhn viscosity, $\eta_K$, and the thermal excitation energy, $k_BT$ for the specific target polymer.

\subsection{Commodity polymer melts at the Kuhn scale}\label{sec:commodity}

\begin{table*}
{
\setlength\tabcolsep{2pt}
\hspace*{-1.0cm}%
\begin{tabular}{|c||c|c|c|c|c|c||c|c|c|c|c||c|c|c|}
\hline 
name
 & $\begin{array}{c} T_{ref}                        \\ \mbox{K}                              \end{array}$
 & $\begin{array}{c} \frac{\langle R^2\rangle}{M_c} \\ \frac{\mbox{\r{A}$^2$mol}}{\mbox{g}}  \end{array}$
 & $\begin{array}{c} \rho_{bulk}                    \\ \frac{\mbox{g}}{\mbox{cm$^3$}}        \end{array}$
 & $\begin{array}{c}  G_e                           \\ \mbox{MPa}                            \end{array}$
 & $\begin{array}{c} p                              \\ \mbox{\r{A}}                          \end{array}$
 & $\begin{array}{c} d_{T}                          \\ \mbox{\r{A}} \end{array}$

 & $\begin{array}{c} n_{K}                          \\ \mbox{}                              \end{array}$
 & $\begin{array}{c} l_{K}                          \\ \mbox{\r{A}}                          \end{array}$
 & $\begin{array}{c} M_{K}                          \\ \frac{\mbox{g}}{\mbox{mol}}           \end{array}$
 & $\begin{array}{c} \frac{M_K}{M_m}                \\ \mbox{} \end{array}$
 & $\begin{array}{c} \rho_{K}               \\ \mbox{nm$^{-3}$} \end{array}$

 & $\begin{array}{c} \frac{G_el_K^3}{k_BT}  \\ \mbox{} \end{array}$
 & $\begin{array}{c} N_{eK}                 \\ \mbox{}\end{array}$
 & $\begin{array}{c} \alpha                 \\ \mbox{} \end{array}$
 \tabularnewline
\hline 
\hline 
%                                                                  lk
PI-50	& 298	& 0.528	& 0.893	& 0.51	& 3.52	& 47.7	& 2.50	& 8.80	& 146.60	& 2.15	& 3.66	& 0.085	& 29.41	& 13.6	\\ \hline
PI-7    & 298	& 0.596	& 0.900	& 0.44	& 3.10	& 55.1	& 2.72	& 8.44	& 119.60	& 1.76	& 4.52	& 0.064	& 42.55	& 17.8	\\ \hline
PDMS$^*$& 298	& 0.422	& 0.970	& 0.25	& 4.06	& 63.7	& 2.82	& 11.42	& 309.28	& 4.17	& 1.89	& 0.091	& 31.08	& 15.7	\\ \hline
PI-20	& 298	& 0.591	& 0.898	& 0.44	& 3.13	& 54.8	& 2.86	& 8.98	& 136.50	& 2.00	& 3.95	& 0.077	& 37.17	& 17.5	\\ \hline
PI-34	& 298	& 0.585	& 0.965	& 0.44	& 2.94	& 56.5	& 3.02	& 9.58	& 156.90	& 2.30	& 3.44	& 0.093	& 32.32	& 19.2	\\ \hline
cis-PBd	& 298	& 0.758	& 0.900	& 0.95	& 2.43	& 42.2	& 3.40	& 8.28	& 90.50 	& 1.67	& 5.99	& 0.131	& 25.93	& 17.3	\\ \hline
PIB(413)& 413	& 0.557	& 0.849	& 0.38	& 3.51	& 65.8	& 3.47	& 12.20	& 267.90	& 4.77	& 1.91	& 0.119	& 29.02	& 18.7	\\ \hline
cis-PI	& 298	& 0.679	& 0.910	& 0.72	& 2.69	& 46.0	& 3.47	& 9.34	& 128.60	& 1.89	& 4.26	& 0.144	& 24.15	& 17.1	\\ \hline
a-PP(463)& 463	& 0.678	& 0.765	& 0.53	& 3.20	& 61.7	& 3.53	& 11.20	& 183.40	& 4.36	& 2.51	& 0.115	& 30.59	& 19.3	\\ \hline
i-PP	& 463	& 0.694	& 0.766	& 0.54	& 3.12	& 61.7	& 3.64	& 11.40	& 187.80	& 4.46	& 2.46	& 0.125	& 29.22	& 19.8	\\ \hline
a-PP(413)& 413	& 0.678	& 0.791	& 0.59	& 3.10	& 56.0	& 3.65	& 11.20	& 183.40	& 4.36	& 2.60	& 0.145	& 25.21	& 18.1	\\ \hline
a-PP(348)& 348	& 0.678	& 0.825	& 0.60	& 2.97	& 51.9	& 3.81	& 11.20	& 183.40	& 4.36	& 2.71	& 0.175	& 21.70	& 17.5	\\ \hline
a-PP	& 298	& 0.678	& 0.852	& 0.60	& 2.87	& 48.8	& 3.92	& 11.20	& 183.40	& 4.36	& 2.79	& 0.205	& 19.15	& 17.0	\\ \hline
PIB	& 298	& 0.570	& 0.918	& 0.43	& 3.17	& 55.2	& 3.94	& 12.50	& 274.20	& 4.89	& 2.02	& 0.202	& 19.52	& 17.4	\\ \hline
a-PMMA	& 413	& 0.390	& 1.130	& 0.39	& 3.77	& 62.5	& 4.07	& 15.30	& 598.00	& 5.97	& 1.14	& 0.243	& 16.72	& 16.6	\\ \hline
i-PS$^*$& 413	& 0.420	& 0.969	& 0.24	& 4.08	& 76.7	& 4.19	& 17.11	& 697.12	& 6.69	& 0.84	& 0.209	& 20.10	& 18.8	\\ \hline
a-PMA	& 298	& 0.436	& 1.110	& 0.31	& 3.43	& 61.9	& 4.29	& 14.70	& 494.60	& 5.75	& 1.35	& 0.241	& 17.79	& 18.1	\\ \hline
PI-75	& 298	& 0.563	& 0.890	& 0.46	& 3.31	& 51.8	& 4.53	& 15.00	& 399.30	& 5.86	& 1.34	& 0.379	& 11.94	& 15.6	\\ \hline
PBd-20	& 298	& 0.841	& 0.895	& 1.34	& 2.21	& 37.3	& 4.54	& 10.10	& 122.40	& 2.26	& 4.41	& 0.335	& 13.55	& 16.9	\\ \hline
a-PS$^*$& 413	& 0.437	& 0.969	& 0.25	& 3.92	& 76.3	& 4.54	& 17.80	& 725.34	& 6.96	& 0.80	& 0.247	& 18.35	& 19.4	\\ \hline
PBd-98	& 300	& 0.661	& 0.890	& 0.71	& 2.82	& 45.4	& 4.83	& 13.70	& 284.80	& 5.27	& 1.88	& 0.442	& 10.93	& 16.1	\\ \hline
PEO$^*$	& 353	& 0.805	& 1.060	& 2.25	& 1.95	& 33.4	& 4.99	& 9.71	& 117.12	& 2.66	& 5.45	& 0.423	& 11.81	& 17.1	\\ \hline
POM$^*$	& 473	& 0.763	& 1.140	& 2.12	& 1.91	& 40.1	& 5.06	& 9.65	& 122.11	& 4.07	& 5.62	& 0.293	& 17.28	& 21.0	\\ \hline
a-PHMA	& 373	& 0.366	& 0.960	& 0.11	& 4.73	& 98.4	& 5.19	& 24.40	& 1622.00	& 9.53	& 0.36	& 0.317	& 16.35	& 20.8	\\ \hline
a-PVA$^*$& 333	& 0.490	& 1.080	& 0.44	& 3.14	& 57.9	& 5.26	& 16.50	& 555.70	& 6.45	& 1.17	& 0.428	& 12.30	& 18.4	\\ \hline
SBR	& 298	& 0.818	& 0.913	& 0.98	& 2.22	& 43.6	& 5.33	& 11.90	& 173.60	& 2.61	& 3.16	& 0.399	& 13.35	& 19.6	\\ \hline
P6N$^*$	& 543	& 0.853	& 0.985	& 2.25	& 1.98	& 41.1	& 5.53	& 10.93	& 140.05	& 1.24	& 4.24	& 0.392	& 14.11	& 20.8	\\ \hline
a-P$\alpha$MS$^*$	& 473	& 0.442	& 1.040	& 0.40	& 3.61	& 67.2	& 5.66	& 20.43	& 944.61	& 7.99	& 0.66	& 0.523	& 10.82	& 18.6	\\ \hline
a-PEA	& 298	& 0.463	& 1.130	& 0.45	& 3.17	& 53.7	& 5.70	& 18.10	& 710.10	& 7.09	& 0.96	& 0.649	& 8.79	& 16.9	\\ \hline
PET$^*$	& 548	& 0.845	& 0.989	& 3.88	& 1.99	& 31.3	& 7.50	& 14.91	& 263.15	& 1.37	& 2.26	& 1.698	& 4.42	& 15.8	\\ \hline
s-PP	& 463	& 1.030	& 0.766	& 1.69	& 2.10	& 42.4	& 7.99	& 16.90	& 278.70	& 6.62	& 1.66	& 1.274	& 6.27	& 20.2	\\ \hline
PE(413)	& 413	& 1.250	& 0.785	& 3.25	& 1.69	& 32.2	& 8.09	& 13.70	& 150.40	& 5.36	& 3.14	& 1.466	& 5.52	& 19.0	\\ \hline
a-POA	& 298	& 0.442	& 0.980	& 0.20	& 3.83	& 73.3	& 8.34	& 31.90	& 2295.00	& 12.45	& 0.26	& 1.578	& 5.29	& 19.1	\\ \hline
PC$^*$	& 473	& 0.864	& 1.140	& 3.38	& 1.69	& 33.9	& 10.93	& 18.43	& 393.25	& 1.55	& 1.75	& 3.237	& 3.38	& 20.1	\\ \hline
PE	& 298	& 1.400	& 0.851	& 4.38	& 1.39	& 26.0	& 11.10	& 15.40	& 168.30	& 6.00	& 3.04	& 3.884	& 2.86	& 18.6	\\ \hline
PTFE$^*$& 653	& 0.598	& 1.460	& 2.12	& 1.90	& 47.2	& 12.30	& 23.40	& 915.41	& 9.15	& 0.96	& 3.019	& 4.07	& 24.8	\\ \hline
\end{tabular}
}

\caption{\label{tab:Kuhn}Kuhn descriptions and entanglement properties of polymers\cite{fetters2007chain}
characterized by experimental temperature $T_{ref}$, mean-square extension per molecular mass $\langle R^2\rangle/M_c$, bulk mass density $\rho_{bulk}$, entanglement modulus $G_e$, and derived quantities such as the packing length $p$ and tube diameter $d_T$. 
The polymers are also characterized by their Kuhn number $n_{K}$, Kuhn length $l_{K}$, mass of a Kuhn segment $M_K$, Kuhn density $\rho_K$, and
the reduced entanglement modulus $G_e l_K^3/[k_BT_{ref}]$,
Kuhn segments between entanglements $N_{eK}$,
and number of entanglement strands per entanglement volume $\alpha$. 
$*$ denotes polymers where we have derived the Kuhn scale descriptions, for the
rest we use the values from Ref. \cite{fetters2007chain}.
To uniquely identify polymers, we have added the reference temperature to some
of the polymer names.}
\end{table*}

At a given state point (temperature), a melt of monodisperse chains (with
molecular weight $M_c$) can be characterized by just a few experimental observables:
the mass density $\rho_{bulk}$, the average chain end-to-end distance per unit mass
$\langle R^{2}\rangle/M_c$, and  the maximal chain extension, $L$.
Values for these observables for a large number of typical polymers are collected in Ref. \cite{fetters2007chain}. We present data for a selected subset of polymers expressed in Kuhn units in Tab. \ref{tab:Kuhn}. The Kuhn lengths are in the $1-2$ nanometer range, with a Kuhn segment mass $M_{K}=M_{c}/N_{K}\sim100-2000$ g/mol. The number of monomers in a Kuhn segment varies in the range of $1-13$, and the number density of Kuhn segments, $\rho_K$, varies in the range of $0.5-5 nm^{-3}$.
%The packing length is derived from the bulk density and end-to-end distance, while the entanglement modulus and bulk density
% provides an %estimate for the tube diameter. While these two observables varies from polymer to polymer, their ratio is
% roughly constant for all flexible polymers.\cite{lin87,kavassalis87}

A key characteristic of polymer species is their Kuhn number, which varies for common, flexible commodity polymers in the range $2 \leq n_K \leq 12$. For comparison,  $n_{K}>10^4$ in gels of tightly entangled filamentous proteins such as f-actin.\cite{Uchida_jcp_08} 
%We observe no systematic correlations between the Kuhn properties of a polymer and the Kuhn
% number, because they depend on chemistry specific details. {\bf I do not understand this
% sentence. What are ``Kuhn properties''? One cannot compare, say, Kuhn times to a dimensionless
%number. To make sense, one has to look at dimensionless numbers like plateau moduli {\em
% expressed in Kuhn units}!} 
In agreement with the arguments in the preceding section, we observe in Tab. \ref{tab:Kuhn} a systematic correlation between the Kuhn number and emergent properties such as the entanglement modulus and the entanglement length measured in Kuhn units. We will return to this point in Sec.~\ref{sec:plateau moduli}.

\subsection{Kremer-Grest model polymer melts at the Kuhn scale}
\label{sec:KGparameterization}

The Kremer-Grest model\cite{grest1986molecular,kremer1990dynamics} is a de facto
standard model in Molecular Dynamics investigations of generic polymer properties.
The KG model is a bead-spring model, where the bead interact via a Lennard-Jones potential. 
Since we are not interested in studying the {\em emergence} of the glass transition\cite{bennemann1998,buchholz2002cooling,grest2016communication}
% {\bf [recent Gary paper and there must be tons of others? Marc Robbins, Rob Hoy, Joerg Roettler]}
, we employ a version with purely repulsive Weeks-Chandler-Anderson (WCA) interactions (the 12-6 Lennard-Jones potential truncated and shifted to zero at the minimum), 
\begin{equation}
U_{WCA}(r<2^{1/6}\sigma)=4k_BT\left[\left(\frac{\sigma}{r}\right)^{-12}-\left(\frac{\sigma}{r}\right)^{-6}+\frac{1}{4}\right]\ ,
%U_{WCA}(r)=4k_BT\left[\left(\frac{\sigma}{r}\right)^{-12}-\left(\frac{\sigma}{r}\right)^{-6}+\frac{1}{4}\right]\quad\mbox{for}\quad r<2^{1/6}\sigma\ ,
\end{equation}
where $\sigma$ defines the bead diameter and where we have explicitly adopted the standard choice to render the model athermal by setting the energy scale of the WCA interaction equal to $k_BT$.
Bonded beads interact through the finite-extensible-non-linear spring (FENE) potential given by
\begin{equation}
U_{FENE}(r)=-15 k_BT\ \left(\frac{R}{\sigma}\right)^{2}\ln\left[1-\left(\frac{r}{R}\right)^{2}\right]\ .
%U_{FENE}(r)=-\frac{kR^{2}}{2}\ln\left[1-\left(\frac{r}{R}\right)^{2}\right]\ .
\end{equation}

We employ the standard choice of $R=1.5\sigma$ for the distance, where the FENE potential diverges. The average bond length is $l_{b}=0.965\sigma$. The standard choice for the bead density is $\rho_b=0.85\sigma^{-3}$.
Faller and M\"u{}ller-Plathe~\cite{faller1999local,faller2000local,faller2001chain} 
augmented the standard KG model with a bending potential,
\begin{equation}\label{eq:Ubend}
U_{bend}(\Theta)=\kappa_\Theta\left(1-\cos\Theta\right),
\end{equation}
where $\Theta$ denotes the angle between subsequent bonds. In the following we
use the reduced bending energy $\kappa \equiv \kappa_\Theta/[k_BT]$.
To prepare the present study, we have investigated the dependence of the characteristic time and length scales in KG bead-spring polymer melts on the reduced bending energy,\cite{svaneborg2018KGCharacterization}. In particular, we found for the Kuhn length:
%{\bf Please reduce ALL your empirical relations to a REASONABLE number of significant digits.}
%{\bf CS: DONE}
%
\begin{eqnarray}
\label{eq:lkkg}
l_K(\kappa)&=& l_K^{(0)} + \Delta l_K\\
\frac{l_{K}^{(0)}(\kappa)}{\sigma} &=& \frac{l_b}\sigma
 \begin{cases}
  \frac{2 \kappa +e^{-2\kappa }-1}{1-e^{-2 \kappa } (2 \kappa +1)} & \text{if $\kappa \neq 0$} \\
  1 & \text{if $\kappa=0$}
\end{cases}\nonumber\\
\frac{\Delta l_K(\kappa)}{\sigma} &=& 
0.77 \left(\tanh \left(-0.03 \kappa^2-0.41 \kappa+0.16\right)+1\right)\nonumber
% 0.769888 + 0.776514 \times\nonumber\\
%&&\tanh \left(0.147244-0.405299 \kappa-0.0298337 \kappa^2\right)\nonumber
\end{eqnarray}
From this relation, we can directly infer the dimensionless Kuhn number, Eq. (\ref{eq:nkdef}), characterising KG melts: 
\begin{equation}\label{eq:nkkg}
n_K(\kappa)= \rho_b \frac{l_b}{l_K}l_K^3\ .
\end{equation}
Finally, the number of Kuhn segments between entanglements, the Kuhn friction and Kuhn time of the KG model are given by
\begin{eqnarray}
N_{eK}(\kappa) &=& -0.84 \kappa^4+3.14 \kappa^3+3.69 \kappa^2-30.1 \kappa+39.3\label{eq:nekinterpolation}\\
%
%N_{eK}(\kappa) &= & 39.130-30.237\kappa + 4.2834\kappa^2 \nonumber\\
%&&  + 3.2065\kappa^3- 1.2879\kappa^4+ 0.1372\kappa^5\label{eq:nekinterpolation}\\
\frac{\zeta{}_{K}(\kappa)}{(m_{b}/\tau)} &=& 
12.8 \left(\frac{l_K(\kappa)}{\sigma} \right)\\
\frac{\tau_{K}(\kappa)}{\tau} &=& 0.434 \left( \frac{l_K(\kappa)}{\sigma} \right)^3
\end{eqnarray}

%
%and from the Kuhn frictions, we derived the following parameterization for the
%Kuhn times
%%
%\begin{equation}\label{eq:taukinterpolation}
%\frac{\tau_{K}(\kappa)}{\tau}
%=4.20761
%+2.8451\frac{\kappa}{\epsilon}
%+1.64852\frac{\kappa^2}{\epsilon^2}
%+0.64643\frac{\kappa^3}{\epsilon^3}
%+0.652419\frac{\kappa^4}{\epsilon^4},
%\end{equation}
%%
%for convenience we also state a parameterization of the entanglement times as
%%
%\begin{equation}\label{eq:taueinterpolation}
%\frac{\tau_{e}(\kappa)}{\tau}
%=6391.76 
%-5723.70\frac{\kappa}{\epsilon}
%+1720.28\frac{\kappa^2}{\epsilon^2}
%-32.19\frac{\kappa^2}{\epsilon^2}
%-42.32\frac{\kappa^3}{\epsilon^3}.
%\end{equation}

The parameterization of the Kuhn length we believe to be valid for arbitrary
values for stiffness $\kappa$, while the other relations hold for bending rigidities
in the interval $-1<\kappa<2.5$. A negative chain stiffness partly
counteracts the stiffness induced by excluded volume interactions between
next-nearest beads along a chain, and hence makes the chain more flexible than
the standard KG model.

\subsection{A one parameter Kremer-Grest ``force field'' for commodity polymer melts}\label{sec:mappingstatic}

\begin{figure}
\includegraphics[angle=270,width=0.5\columnwidth]{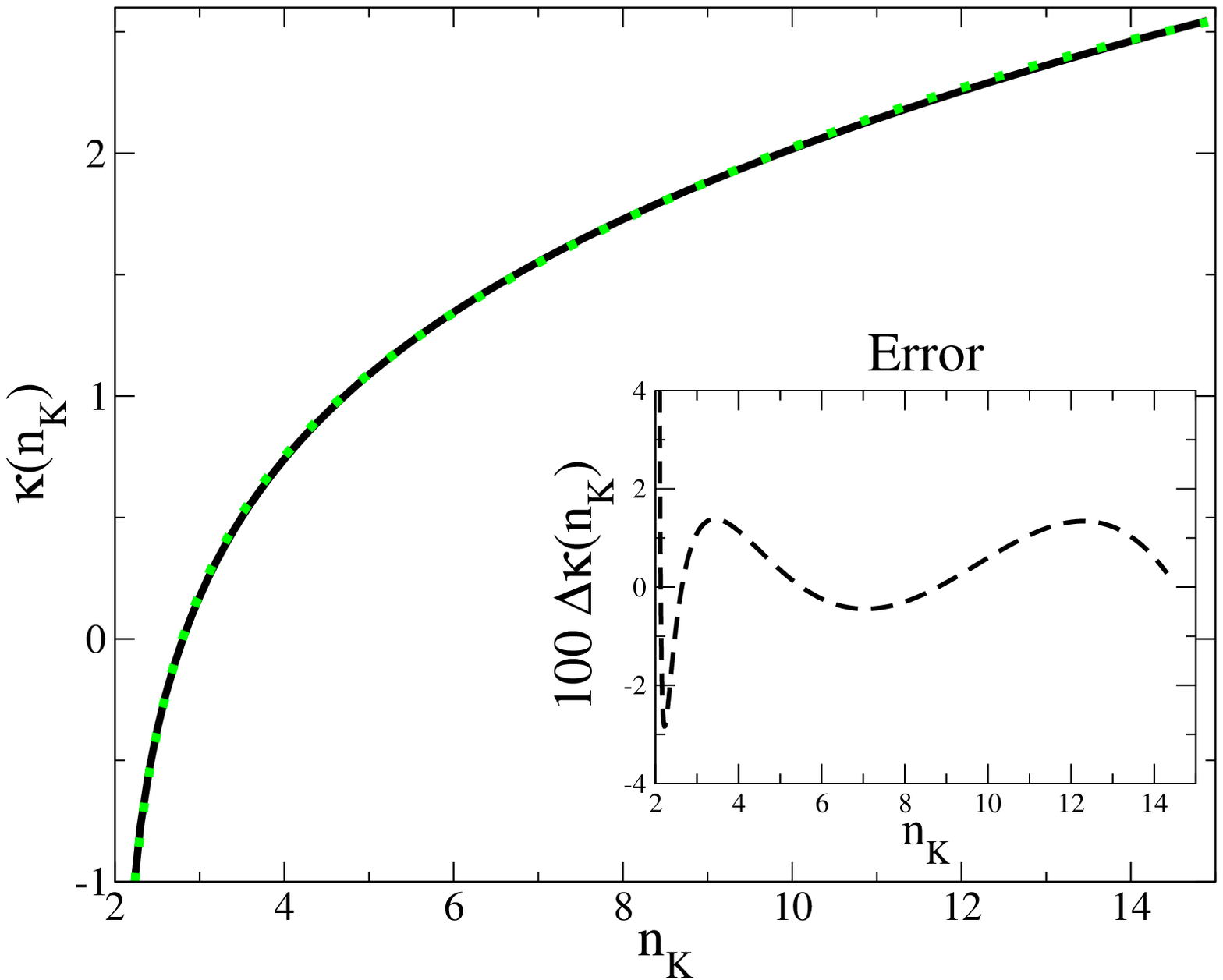}
\caption{\label{fig:kappafit}KG chain stiffness vs. Kuhn number
using eq. (\protect\ref{eq:nkkg}) (solid black line) and our
approximate inversion eq. (\protect\ref{eq:kappafromnk}) (green symbols).
The inset shows the error of our numerical inversion.
}
\end{figure}

%As a first step in defining a KG force field for a polymeric material we fix the energy scale to the thermal excitation energy~\cite{kroeger2004simple},
%%
%\begin{equation}\label{eq:epsilon}
%\epsilon=k_B T\ ,
%\end{equation}
%at the temperature of interest.
%The key step is the choice of the bending stiffness: 

To define a KG force field for a polymeric material we match the dimensionless Kuhn numbers $n_K$ characterising
the experimental system and the model polymer melt. A priori, this requires the numerical inversion of the
combination of Eqs.~ (\ref{eq:lkkg}) and  (\ref{eq:nkkg}) to identify the corresponding reduced stiffness to
use in the KG model. As shown in Fig.~\ref{fig:kappafit},  the approximate relation
%
%\begin{eqnarray}\label{eq:kappafromnk}%UPDATED
%\kappa(n_K)= && 0.822264 \log (n_K-2.0)+0.280663\nonumber\\
%&&-0.0552417 n_K+0.00871154 n_K^2-0.000290834 n_K^3
%\kappa(n_K)= && 0.824 \log (n_K-2.0)+0.28\nonumber\\
%&&-0.055 n_K +0.0087n_K^2-0.00029 n_K^3
%\end{eqnarray}
\begin{equation}\label{eq:kappafromnk}%UPDATED
%\kappa(n_K)= && 0.822264 \log (n_K-2.0)+0.280663\nonumber\\
%&&-0.0552417 n_K+0.00871154 n_K^2-0.000290834 n_K^3
\kappa(n_K)=  0.824 \log (n_K-2.0)  -0.00029 n_K^3  +0.0087n_K^2    -0.055 n_K     +0.28  
\end{equation}
provides an excellent approximation over the experimentally relevant range, 
$2 \le n_K \le 15$. Through eq. (\ref{eq:kappafromnk}), $l_K$, $N_{eK}$, $\zeta_K$, and $\tau_K$
become functions of the Kuhn number. Note that the standard KG model with $\kappa=0$
essentially corresponds to the intrinsically most flexible polymers such as PDMS or PI
with $7$ to $50\%$ 3,4 content.

\begin{table*}

\begin{tabular}{|c|c||c|c|c|c|c||c|c|c|}
\hline 
name
 & $n_K$
 & $\kappa$
 & $c_{b}$
 & $\frac{l_{K}}{\sigma}$
 & $\frac{M{}_{m}}{M_{b}}$ 
 & $\frac{M_{b}}{\mbox{[g/mol]}}$ 
 & $\frac{k_B T_{ref}}{10^{-21}\mbox{J}}$
 & $\frac{\sigma}{\mbox{nm}}$ 
 & $\frac{k_B T_{ref}\sigma^{-3}}{\mbox{MPa}}$ 
 \tabularnewline
\hline 
\hline
%                 nk      kappa           cb      lk     mm/mb      Mb            eps   sigma    press/MPa
PI-50	        & 2.50	& -0.378	& 1.81	& 1.74	& 0.84	& 81.13 	& 4.11  & 0.50	& 32.0	\\ \hline
PI-7	        & 2.72	& -0.086	& 1.89	& 1.82	& 1.07	& 63.37 	& 4.11  & 0.46	& 41.3	\\ \hline
PDMS$^*$	& 2.82	& 0.013	        & 1.92	& 1.85	& 0.46	& 161.05	& 4.11	& 0.62	& 17.6	\\ \hline
%IPDMS$^*$       & 3.04  & 0.205 & 2.00  & 1.93  & 0.48  & 154.97        & 4.11  & 0.59  & 19.7  \\ \hline

PI-20   	& 2.86	& 0.056	        & 1.94	& 1.87	& 0.97	& 70.50 	& 4.11	& 0.48	& 37.1	\\ \hline
PI-34	        & 3.02	& 0.191	        & 1.99	& 1.92	& 0.86	& 78.86	        & 4.11	& 0.50 	& 33.1	\\ \hline
cis-PBd	        & 3.40	& 0.445	        & 2.11	& 2.04	& 1.26	& 42.87 	& 4.11	& 0.41 	& 61.3	\\ \hline
PIB(413)	& 3.47	& 0.483	        & 2.13	& 2.06	& 0.45	& 125.69	& 5.70	& 0.59	& 27.3	\\ \hline
cis-PI	        & 3.47	& 0.484 	& 2.13	& 2.06	& 1.13	& 60.32	        & 4.11	& 0.45 	& 44.0	\\ \hline
%Icis-PI & 3.04  & 0.205 & 2.00  & 1.93  & 1.06  & 64.44 & 4.11  & 0.48  & 36.1  \\ \hline
a-PP(463)	& 3.53	& 0.518	        & 2.15	& 2.07	& 0.49	& 85.25	        & 6.39  & 0.54	& 40.7	\\ \hline
i-PP	        & 3.64	& 0.575 	& 2.18	& 2.11	& 0.49	& 85.96	        & 6.39	& 0.54 	& 40.4	\\ \hline
a-PP(413)	& 3.65	& 0.580	        & 2.19	& 2.11	& 0.50	& 83.84	        & 5.70  & 0.53	& 38.2	\\ \hline
a-PP(348)	& 3.81  & 0.656	        & 2.23	& 2.16	& 0.51	& 82.08	        & 4.80	& 0.52 	& 34.3	\\ \hline
a-PP	        & 3.92	& 0.708	        & 2.27	& 2.19	& 0.52	& 80.84	        & 4.11	& 0.51 	& 30.7	\\ \hline
PIB	        & 3.94	& 0.714	        & 2.27	& 2.19	& 0.47	& 120.66	& 4.11	& 0.57 	& 22.2	\\ \hline
a-PMMA	        & 4.07	& 0.770	        & 2.31	& 2.23	& 0.39	& 258.82	& 5.70	& 0.69 	& 17.6	\\ \hline
i-PS$^*$	& 4.19	& 0.819	        & 2.35	& 2.26	& 0.35	& 297.22	& 5.70	& 0.76	& 13.2	\\ \hline
a-PMA	        & 4.29	& 0.856	        & 2.37	& 2.29	& 0.41	& 208.42	& 4.11	& 0.64 	& 15.6	\\ \hline
PI-75	        & 4.53	& 0.941 	& 2.44	& 2.35	& 0.42	& 163.78	& 4.11	& 0.64 	& 15.9	\\ \hline
PBd-20	        & 4.54	& 0.944 	& 2.44	& 2.35	& 1.08	& 50.16	        & 4.11	& 0.43 	& 52.2	\\ \hline
a-PS$^*$	& 4.54	& 0.944	        & 2.44	& 2.35	& 0.35	& 297.19	& 5.70	& 0.76	& 13.2	\\ \hline
PBd-98	        & 4.83	& 1.039	        & 2.52	& 2.43	& 0.48	& 113.08	& 4.14	& 0.56 	& 23.1	\\ \hline
PEO$^*$	        & 4.99	& 1.086	        & 2.56	& 2.47	& 0.96	& 45.77	        & 4.87	& 0.39 	& 80.2	\\ \hline
POM$^*$	        & 5.06	& 1.105	        & 2.58	& 2.48	& 0.63	& 47.40	        & 6.53	& 0.39 	& 111.5	\\ \hline
a-PHMA	        & 5.19	& 1.143	        & 2.61	& 2.52	& 0.27	& 621.59	& 5.15	& 0.97 	& 5.7	\\ \hline
a-PVA$^*$	& 5.26  & 1.162	        & 2.63	& 2.53	& 0.41	& 211.52	& 4.60	& 0.65	& 16.7	\\ \hline
SBR	        & 5.33	& 1.182	        & 2.65	& 2.55	& 1.01	& 65.62	        & 4.11	& 0.47 	& 40.6	\\ \hline
P6N$^*$	        & 5.53	& 1.234	        & 2.69	& 2.60	& 2.18	& 51.98	        & 7.50	& 0.42 	& 100.9	\\ \hline
a-P$\alpha$MS$^*$  & 5.66 & 1.265	& 2.72	& 2.63	& 0.34	& 346.66	& 6.53	& 0.78	& 13.9	\\ \hline
a-PEA	        & 5.70	& 1.276	        & 2.74	& 2.64	& 0.39	& 259.56	& 4.11	& 0.69  & 12.8	\\ \hline
PET$^*$	        & 7.50	& 1.646	        & 3.14	& 3.03	& 2.29	& 83.82	        & 7.57	& 0.49  & 63.4	\\ \hline
s-PP	        & 7.99	& 1.728	        & 3.24	& 3.12	& 0.49	& 86.03	        & 6.39	& 0.54  & 40.5	\\ \hline
PE(413)	        & 8.09	& 1.744	        & 3.26	& 3.14	& 0.61	& 46.15	        & 5.70	& 0.44  & 69.0	\\ \hline
a-POA   	& 8.34	& 1.785	        & 3.31	& 3.19	& 0.27	& 693.22	& 4.11	& 1.00  & 4.1	\\ \hline
PC$^*$	        & 10.93	& 2.136	        & 3.79	& 3.65	& 2.45	& 103.76	& 6.53	& 0.50  & 51.0	\\ \hline
PE	        & 11.10	& 2.156	        & 3.82	& 3.68	& 0.64	& 44.07	        & 4.11	& 0.42 	& 56.4	\\ \hline
PTFE$^*$	& 12.30	& 2.291	        & 4.02	& 3.87	& 0.44	& 227.70	& 9.02	& 0.60 	& 41.1	\\ \hline
\end{tabular}

\caption{\label{tab:Kremer-Grest-model-parameters}Kremer-Grest model parameters
for the polymers shown in Tab. \ref{tab:Kuhn} in terms of
the Kuhn number, bending stiffness $\kappa$, number of beads per Kuhn
segment $c_b$, Kuhn length expressed in KG units, number of beads per monomer $M_m/m_b$,
and finally the conversion relations from KG units for energy $k_B T_{ref}$, length $\sigma$, and stress $k_B T_{ref}\sigma^{-3}$ to SI units. 
}
\end{table*}

The number of beads per Kuhn length is given by
\begin{equation}\label{eq:beads per lK}
c_b(n_K) \equiv \frac{l_K}{l_b} = \sqrt{\frac{n_K}{\rho_bl_b^3}},
\end{equation}
and hence the number of beads per chain required to model an experimental
target polymer melt with chain length $N_K$ is
\begin{equation}\label{eq:beads per chain}
N_b=c_b(n_K)N_K\ .
\end{equation}

What remains is to fix the mapping relations for the simulation units of length,
mass, and time. Equating the model and experimental Kuhn lengths and accounting for the small difference, $l_b=0.965\sigma$, between the bond length and the bead diameter in the KG model, we obtain
\begin{equation}
\sigma = \frac{l_K^{exp}}{0.965\times c_b(n_K)}.
\end{equation}

The bead mass is obtained along the same lines by equating the experimental mass
of a Kuhn segment to the mass of a Kuhn segment in the model:
\begin{equation}
m_b = \frac{M_K^{exp}}{c_b(n_K)}\ .
\end{equation} 

In Table~\ref{tab:Kremer-Grest-model-parameters} we have listed the resulting Kremer-Grest model parameters and mappings for  the polymer species shown in Tab. \ref{tab:Kuhn}. By construction, the number of beads per Kuhn length is an increasing function of $n_K$ and varies between 1.7 and 3.9. With $-0.4\le \kappa \le 2.3$ the required stiffness parameters falls into the validity range of our empirical relations for Kuhn length, entanglement length, and Kuhn friction. 
Bead diameters vary between $4$ and $10\mathring{A}$, the energy scale is given by the experimental temperature range and varies within a factor of two. Nevertheless,  the KG unit of stress, $4  M Pa< k_B T\sigma^{-3} < 112 M Pa$, exhibits a much larger spread. 
As a rule of thumb, beads correspond to monomers. But there are important variations. To cite some examples, 
\begin{description}
\item[PI and PDMS] (polyisoprene and polydimethylsiloxane) are effectively the most flexible chains, which map fairly well on the standard KG model with $\kappa\equiv0$. PI beads have a diameter of $5\mathring{A}$ and represent one monomer; PDMS beads have a diameter of $6\mathring{A}$ and represent two monomers. 
\item[PS] (polystyrene) beads represent three monomers and have with $7.6\mathring{A}$ a correspondingly larger diameter.
\item[PE] (polyethylene) is among the effectively stiffest chains, which 4 beads per Kuhn length and $\kappa\approx2$. PE beads represent 1.5 monomers; with $4\mathring{A}$ they are relatively small. 
\item[PC] (polycarbonate) is comparable to PE in effective stiffness. With a diameter of $5\mathring{A}$, PC beads are comparable to PI beads. However, in the case of PC 2.5 beads are required to represent the more complex monomers, which is remarkably similar to 2 bead / ellipsoid models per monomer used by Tschoeb et al.\cite{tschop1998simulation}
%
% Cameron and later people used a 4:1 CG scheme. 
%
\end{description}
Note that while we provide force-fields for materials like polyethylene (PE), polyoxymethylene (POM), polyethylene terephthalate (PET), polybutylene terephthalate (PBT), polytetrafluoroethylene (PTFE) or isotactic polypropylene (i-PP), they cannot be expected to reproduce the tendency to form semi-crystalline ordering. Such KG models should thus be taken with a grain of salt or, maybe, as a reminder that there is more to polymers than universal properties. Nonetheless, we note that coarse-grain models have been used to study crystallization\cite{meyer2001formation}, and recently specialized KG models were developed and optimized\cite{hoy2013simple,morthomas2017crystallization} to study crystallization phenomena.

\subsection{Temperature dependence of the parameters}\label{sec:temperature dependence static mapping}
{\em A priori}, the parameters listed in Table~\ref{tab:Kremer-Grest-model-parameters} are only valid at the indicated reference temperatures, where the chain dimensions were determined experimentally. 
%What happens if we want to model melts at a different temperature? 
To model polymer melts at different temperatures, we have, in principle, to account for changes in (i) the single chain statistics and (ii) the overall density. 
While it is straightforward to rationalize~\cite{MEYER20131904} a temperature dependent Kuhn length, temperature variations in $n_K$ due to density changes result in less intuitive shifts in bead diameters and weights with temperature. This is a consequence of our choice to preserve the ``canonical'' KG bead density of $\rho_b=0.85 \sigma^{-3}$.

%It is straightforward to translate a temperature dependent Kuhn length, $l_K = l_K(T)$, into a temperature dependent bending {\em free} energy, $\kappa = \kappa(T)$, of our coarse-grain model. To a first approximation,  $\kappa(T)$ can be written in the form $\kappa(T) = \kappa_h - \kappa_s T$, where $\kappa_h$ and $\kappa_s$ are the enthalpic and entropic contributions to the stiffness.\cite{MEYER20131904} Since the Kuhn length of the KG model, Eq.~(\ref{eq:lkkg}), depends on the {\em reduced} bending free energy, $x = \kappa(T)/k_BT$, a temperature-{\em independent} Kuhn length corresponds to a bending free energy of {\em entropic} origin.
%Note, however, that density changes and the corresponding temperature variations in $n_K$ result in less intuitive shifts in bead diameters and weights with temperature. This is a consequence of our choice to preserve the ``canonical'' KG bead density of $\rho_b=0.85 \sigma^{-3}$.
%
%%ZQEX: MUCH BETTER

In practice, the static melt properties are relatively insensitive to changes of temperature: the relative density expansion coefficient is $d\ln\rho_{bulk}/dT \approx -6\times 10^{-4} K^{-1}$, while typical thermal chain expansion coefficients $|d\ln\langle R^2\rangle(T)/dT| < 10^{-3}K^{-1}$.\cite{fetters2007chain} Since $n_K(T) \sim \rho_{bulk} \langle R^2\rangle^2$, we obtain $|d\ln n_K(T)/dT| < 3\times 10^{-3}K^{-1}$. 
In the one case where Ref.~\cite{fetters2007chain} provides data, atactic polypropylene, the 50\% increase in temperature over the interval $298 K\le T \le463K$ causes a slight increase in density while apparently leaving the chain dimensions unchanged. The corresponding reduction of the Kuhn number from $n_K=3.92$  to $n_K=3.43$, suggests that one changes the bead weights from $81$ to $85$ $g/mol$, the bead diameters change from $5.1$ to $5.4$ $\mathring{A}$, while the required reduction of the bending stiffness decreases the Kuhn length in LJ units from $l_K=2.19\sigma$ to $l_K=2.07\sigma$.
Compared to the dynamic effects discussed in the following section, it thus seems safe to transfer the $\kappa$, $M_b$ and $\sigma$ values listed in Tab.~\ref{tab:Kremer-Grest-model-parameters} to other temperatures. 
%This suggests that the {\em effective} bending rigidity at the KG-level appears is essentially entropic in origin
%%, $\kappa_h \ll  -T_{ref} \kappa_s$, 
%and related to rotations between isomeric states~\cite{flory1969statistical}.
Compared to the standard "one-model-fits-all commodity polymer melts" approach discussed in the introduction, we thus suggest the use of chemistry-specific athermal models over the entire (not extremely wide) experimentally relevant temperature range. 
While the use of {\em entropic} springs is standard in coarse-grain polymer models since the earliest theories of rubber elasticity~\cite{TreloarBook}, an entropic wormlike bending rigidity like in Eq.~(\ref{eq:Ubend}) might appear unusual. An alternative, could be KG models with freely rotating bonds along the lines of Ref.~\cite{HsuKremer2019}. However, within the present ansatz the resulting behaviour is described by an athermal bending term.
Obviously, the relations provided above can be used to obtain an improved parameterization, if there is information available about the end-to-end distance and bulk density at the state point of interest.

\subsection{Time mapping}\label{sec:mappingtime}

\begin{table*}
{
\setlength\tabcolsep{2pt}
\hspace*{-2.0cm}%
\begin{tabular}{|c|| c|c|c|| c|c|c|c|| c|c||}
\hline 
name & 
$T_{g} [K]$ & $T_{ref}^{dyn} [K]$ & $\tau_e^{exp} [s]$  &
$n_K$ & $N_{eK}$ & $\tau_K(T_{ref}^{dyn}) [ns] $ & $\eta_K(T_{ref}^{dyn}) [mPa\ s]$ &
$\tau(T_{ref}^{dyn}) [ns]$   & $\sigma \sqrt{M_b/k_BT_{ref}^{dyn}} [ps]$
\tabularnewline
\hline 
%
%PI-X 1,4-polyisoprene where x = 3,4 content;
%
PI-7 & 
$206$ & $298$  & $ 1.9\times 10^{-5}$ &
$2.72$ & $41.76$ & $10.90$ & $61.32$ &
$4.16$ & $2.33$
\tabularnewline

\hline 
PDMS & $150$  & $298$ &  $1.1\times10^{-7}$ &
$2.82$ & $38.74$ & $0.073$ & $0.17$ &
$0.027$ & $5.00$
\tabularnewline

\hline 
cis-PBd & $174$ & $298$ & $8.8\times 10^{-8}$ &
$3.40$ & $26.76$ & $0.12$ & $0.73$ & $0.034$ & $1.71$
\tabularnewline

\hline 
cis-PI & $206$ & $298$ &  $6.7\times 10^{-5}$  &
$3.47$ & $25.80$ & $100.69$ & $418$ &
$26.69$ & $2.22$
\tabularnewline

\hline
a-PP  & $262$ & $348$  & $ 1.9\times10^{-6}$ &
$3.81$ & $21.82$ & $3.99$ & $11.22$ &
 $0.92$ & $2.77$
\tabularnewline

\hline 
PIB  & $201$ & $298$ & $1.1\times 10^{-3}$  &
$3.94$ & $20.58$ & $2697$ & $4499$ &
 $568.8$ & $3.98$
\tabularnewline

\hline 
a-PS    &  $375$ &  $453$ & $3.4\times10^{-4}$  &
$4.54$ & $16.18$ & $1299$ & $1184$ &
 $229.7$ & $6.75$
\tabularnewline

\hline 
PEO    & $210$ &  $348$ & $1.5\times10^{-8}$  &
$4.99$ & $13.87$ & $0.078$ & $0.34$ &
 $0.012$  & $1.55$
\tabularnewline

\hline
\end{tabular}
}
\caption{\label{tab:PhysicalPropertiesTime}
Parameters and characteristic times for commodity polymer melts and the corresponding KG models.
The first set of numbers defines the experimental input: the experimental glass transition temperature and the dynamic reference temperature $T_{ref}^{dyn}$ for the experimental entanglement time $\tau_e$ (and/or more suitable VF parameters).
Using the static mapping and, in particular, the Kuhn number $n_K$, we can infer the number of Kuhn segments per entanglement length, $N_{eK}$, from Eq.~(\ref{eq:netheory}) or more refined estimates~\cite{uchida2008viscoelasticity,svaneborg2018KGCharacterization}. The Kuhn time, $\tau_K(T_{ref}^{dyn})$, and the viscosity at the Kuhn scale, $\eta_K(T_{ref}^{dyn})$, follow from Eqs. (\ref{eq:tauk}) and (\ref{eq:taue_over_tauK}), the characteristic time scale, $\tau(T_{ref}^{dyn})$, of the corresponding Kuhn mapped KG model is given by Eq.~(\ref{eq:tauk}). For comparison, we also list the estimate for $\tau$ that results from the static mapping.
Finally, we can use Eqs.~(\ref{eq:tts}) and (\ref{eq:VF Kuhn time}) to estimate $\tau_K$, $\eta_K$, and $\tau$ over the entire TTS validity range (Fig.~\ref{fig:tts}).
References for experimental data:
PI-7\cite{Gotro1984,abdel2004rheological},
PDMS\cite{vaca2011time}, cis-PDb\cite{vaca2011time}, cis-PI\cite{vaca2011time},
cis-PI\cite{auhl2008linear},
a-PP\cite{van2004method},
%
%i-PP, PIB \cite{van2004method},
%
%a-PS \cite{likhtman2002quantitative} TTS parameters from \cite{ngai2007temperature},
%
and PEO \cite{niedzwiedz2008chain}.
%
%PE shifted using eq. \ref{eq:arrhenius} with entanglement times and activation energy
%$E_a=25kJ/mol$ given in Ref. \cite{ramos2008entanglement}.
}
\end{table*}

\begin{figure}
\includegraphics[angle=270,width=0.5\columnwidth]{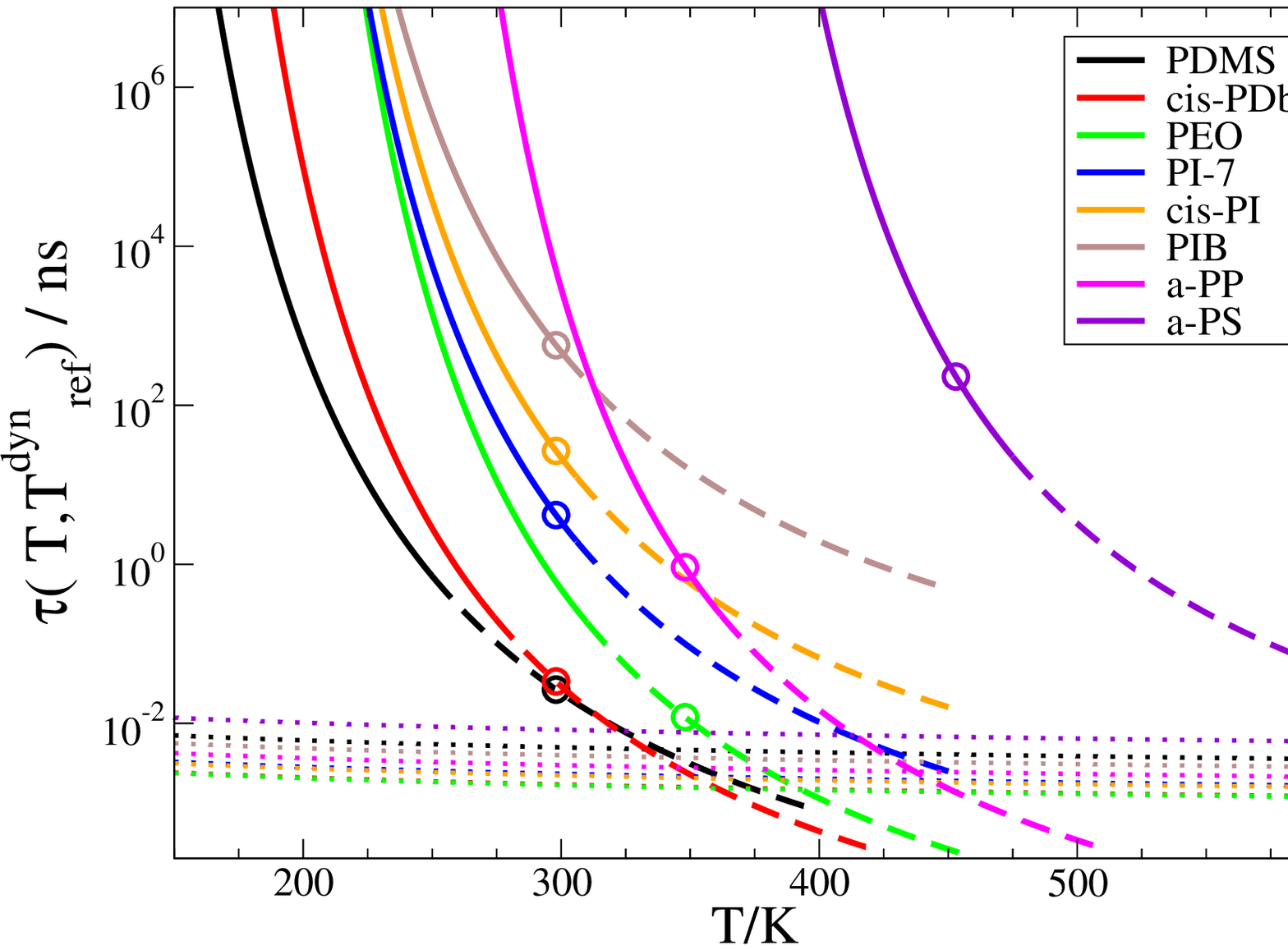}
\caption{\label{fig:tts}
KG time unit $\tau$ in $ns$ as function of temperature for a number of polymer species listed in the legend and distinguished by color.
Typical time steps in simulations are $\delta t = 10^{-2}\tau$.
Solid lines: WLF-extrapolation over the temperature range $[T_g,T_g+100K]$,
Thick dashed lines: WLF-extrapolation for $T>T_g+100K$,
Symbols: Estimate of $\tau$ derived from experimental data for the dynamic
reference temperature, $T_{ref}^{dyn}$, underlying the WLF extrapolation.
Thin dashed lines: standard estimation of the LJ time $\tau=\sigma \sqrt{M_b/[k_B T_{ref}^{dyn}]}$ using the mapping values for bead diameter, bass, and energy scale.
}
\end{figure}

%Gary paper on standard KG with attraction: https://aip.scitation.org/doi/10.1063/1.4964617

To reproduce not only static but also dynamic properties of target systems, we require input on their Kuhn time, $\tau_K$, or their effective viscosity, $\eta_K$, at the Kuhn scale. Equating with $\tau_K$ or $\eta_K$ of the Kuhn mapped KG model, we can directly infer the value of the KG time unit $\tau$ in SI units from Eq.~(\ref{eq:tauk}), since the value of $\kappa(n_K)$ is known via Eq.~(\ref{eq:kappafromnk}) for the Kuhn number of the experimental system. However, to carry out this program, we needed to overcome two difficulties.

While conceptually useful, $\tau_K$ and $\eta_K$ are not straightforward to observe directly. Typically, one can extrapolate down to the Kuhn scale within a model, if there is information on (emergent) macroscopic behavior or time scales at some dynamic reference temperature, $T_{ref}^{dyn}$. Examples of other observables, that are easier accessible experimentally, and from which we can obtain a time mapping is the viscosity of unentangled chains, $\eta = n_K N_K\  \eta_K$, the entanglement time, $\tau_e=N_{eK}^2\tau_K$, the Rouse time, $\tau_R=N_{K}^2\tau_K$, or the terminal relaxation time, $\tau_{max}$. 
%Examples are the viscosity of unentangled chains, $\eta = n_K N_K\  \eta_K$, or time scales like the entanglement time, $\tau_e(n_K)=N_{eK}(n_K)^2\tau_K(n_K)$, the Rouse time, $\tau_R(n_K)=N_{K}^2\tau_K(n_K)$, or the terminal relaxation time is $\tau_{max}(n_K)$. 
%Simulation data for atomistic models {\bf [references to stuff from FMP, Theodorou and the other Greeks, ....]} could be treated along the lines of our analysis of the KG model~\cite{svaneborg2018KGCharacterization} or simply shifted {\bf FMP?!} is to superimpose 
% 	NOT SURE IF WE SHOULD MENTION THEM AS AN ALTERNATIVE WAY TO PARAMETERIZE
%	WOULD THERE BE ANY ADVANTAGE IN DOING THAT COMPARED TO EXP?????
%
%
Experimentally, the Kuhn time or equivalently the Kuhn friction can be obtained from neutron spin echo data \cite{richter2005neutron} by applying expressions from Rouse theory to analyse the  monomeric dynamics below the entanglement time scale as in our analysis of simulation data~\cite{svaneborg2018KGCharacterization}. The entanglement time, $\tau_e$, can be measured by oscillatory rheological experiments, dielectric relaxation and transverse relaxation NMR measurements, see e.g. Refs.~\cite{Bird_77,Watanabe,mcleish02,SAALWACHTER2007}. 
We note that published estimates might be obtained by fitting data to expressions, which define these times using conventions for prefactors, which differ from those we have adopted here. 
%
%, which are all derived from rheological data analyzed in the framework of the Likhtman-McLeish\cite{LikhtmanMcLeishMM02}
% theory for the sake of consistency.
%
% Most but not all are LM. PDMS\cite{vaca2011time}, cis-PDb\cite{vaca2011time}, cis-PI\cite{vaca2011time} is a VF equation,
% reference is the JE Mark polymer handbook.

%\subsection{Time-temperature superposition}\label{sec:TTS}

The second difficulty is the pronounced temperature dependence of the chain dynamics in polymer melts. Most commodity polymer melts become glassy below a temperature $T_g$ in or slightly below the experimentally relevant temperature range. As a consequence, even a small change in temperature can have a significant impact on the dynamics. 
%{\bf crystallization is another issue. Is that a transition at a well defined temperature? And do people still use some kind of TT superposition in the melt regime?????}
Experimentally, time-temperature superposition (TTS) \cite{ngai2007temperature}
%%
%\begin{equation}\label{eq:tts}
%\tau_{exp}(T)=\tau_{exp}(T_0)10^{\log a_T}
%\end{equation}
is used to explore polymer dynamics over a much wider range of frequencies than those directly accessible to a given measurement instrument. Here we use this approach to estimate the Kuhn time at the temperature of interest, $T$, given a Kuhn time measured at the reference temperature, $T_{ref}^{dyn}$:
\begin{equation}\label{eq:tts}
\frac{\tau_{K}(T)}{\tau_{K}(T_{ref}^{dyn})} 
= a_T(T,T_{ref}^{dyn})
\end{equation}
As discussed in appendix~\ref{sec:TTS}, the shift factor can be written as
\begin{equation}\label{eq:VF Kuhn time}
\ln a_T(T,T_{ref}^{dyn})=-\frac{C_{VF}(T-T_{ref}^{dyn})}{(T_{ref}^{dyn}-T_{VF})(T-T_{VF})}\ .
\end{equation}
Eq.~(\ref{eq:VF Kuhn time}) should be valid above the glass transition temperature in the temperature range $[T_g, T_g+100K]$.
Here we use a ``universal'' Vogel-Fulcher constant $C_{VF} = \ln(10) \times 17.44 \times 51.6K = 2072 K$. Similarly, we set $T_{VF} = T_g-51.6K$ for the Vogel-Fulcher temperature, where viscosities and associated time scales formally diverge. More detailed information and specific tables with fitted VF (or WLF, see appendix~\ref{sec:TTS}) parameters can be found in Ref.~\cite{ngai2007temperature}.

The mapping relations for the temperature dependent conversion of the KG time $\tau$ resulting from Eq.~(\ref{eq:tauk}) are shown in Tab. \ref{tab:PhysicalPropertiesTime} and illustrated in Fig.~\ref{fig:tts}.
%With $7ps \le \tau \le 0.35\mu s$ our estimates are in reasonable agreement with $1\tau \sim 2ps-31ns$  originally estimated by Kremer and Grest\cite{kremer1990dynamics}. {\bf True? Which polymers did they look at? Probably not worth citing their range of times, if it is for different polymers} 
%In particular, 
The converted values vary over a much wider range than the static parameters in Table~\ref{tab:Kremer-Grest-model-parameters}:
\begin{description}
\item[PDMS, cis-PDb and PEO] are experimentally studied about $150K$ above $T_g$, resulting in KG time scales in the $10ps$ range.
%\item[PI-7] ARE YOU SURE THAT YOUR $T_{REF}^{DYN}$ IS CORRECT??? FIXED to 298K
\item[PI] melts at $100K$ above $T_g$ are represented by KG models with $\tau$ in the $10 ns$ range.
\item[PIB] has a significantly higher $\tau\approx 350 ns$ at a similar distance from the glass transition temperature. Perhaps this can be explained by specific intramolecular rotational barriers.\cite{arbe2001origin}.
\item[a-PS] has a comparable $\tau\approx 140 ns$ at $80K$ above $T_g$, while
\item[a-PP] maps onto a KG model with $\tau\approx 0.6 ns$ at a comparable distance from $T_g$
\end{description}

%We observe a huge difference between $\tau$ obtained by matching experimental time scales,
%and from the expression we used to define the unit of time $\sigma\sqrt{M_b/k_BT}$.

%Needless to say, that improved estimates could be obtained by superimposing dynamic data from simulations and experiments for systems of {\em identical} but not necessarily {\em ideal} (i.e. monodisperse) composition.

%Hence we can expect an exponential acceleration of real time vs. one unit of simulation time as the experimental temperature is decreased relative to the reference temperature. 
% we say this in discussion/conclusion together with the obvious downside, that as for static properties this comes at the price of not being able to PREDICT the values of these parameters.

\subsection{A first test: plateau moduli of commodity polymer melts}\label{sec:plateau moduli}

\begin{figure}
\includegraphics[angle=270,width=0.5\columnwidth]{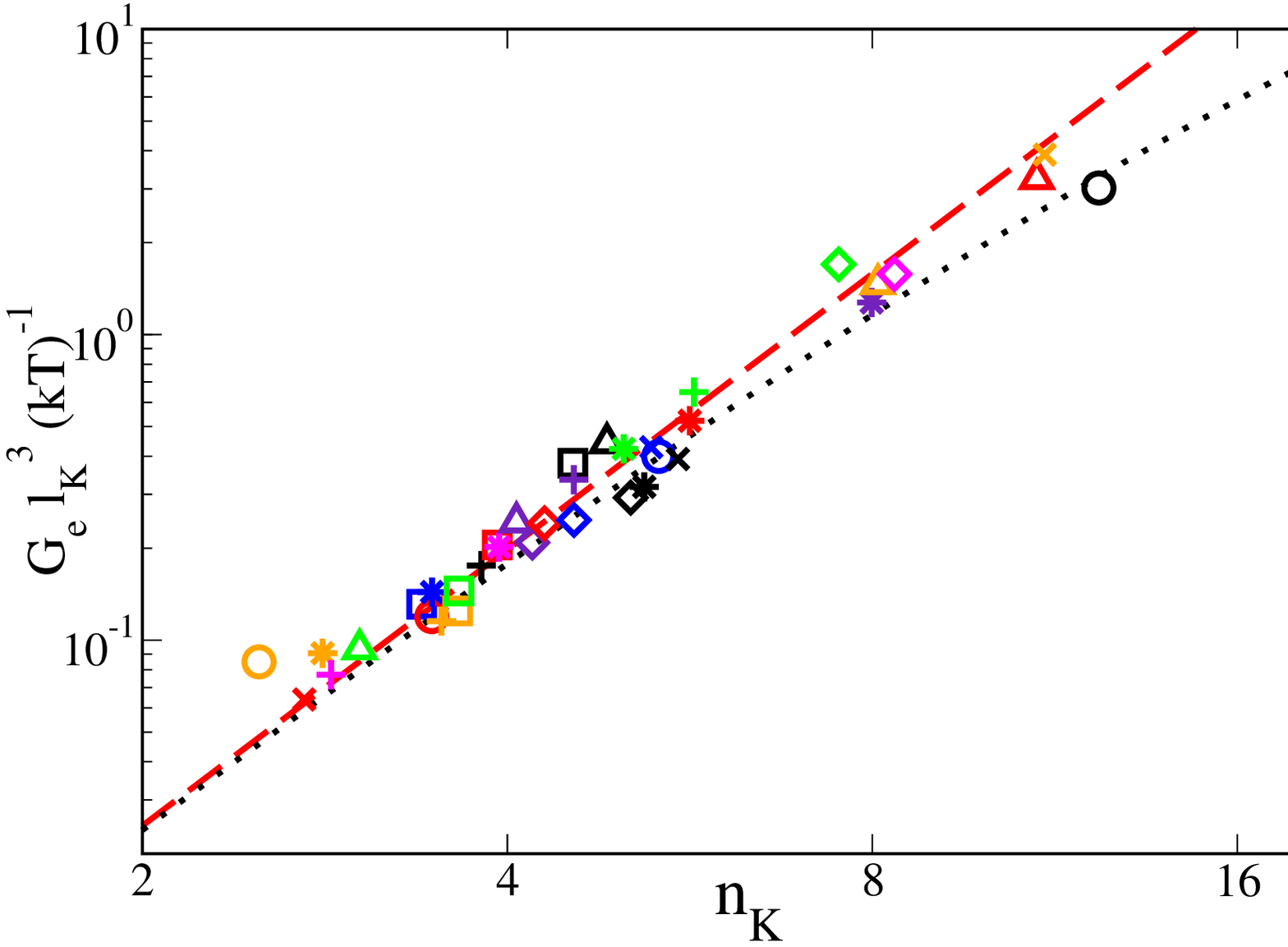}
\caption{\label{fig:reducedplateaumoduli1}Reduced entanglement moduli for the polymers in
Tab. \ref{tab:Kuhn} compared to the theoretical expectation for flexible polymers eq. (\ref{eq:reducedmodulus}) (red dashed line) to the semi-empirical prediction of eq. (42) in Ref. \cite{svaneborg2018KGCharacterization}) (black dotted line)
The symbols denote in order of increasing Kuhn number:
PI-50 (orange $\circ$),
PI-7 (red $\times$),
PDMS (orange $\ast$),
PI-20 (magenta $+$),
PI-34 (green $\bigtriangleup$),
cis-PBd (blue box),
PIB(413) (red $\circ$),
cis-PI (blue $\ast$),
a-PP(463) (orange $+$),
i-PP (orange box),
a-PP(413) (green box),
a-PP(348) (black $+$),
a-PP (red box),
PIB (magenta $\ast$),
a-PMMA (indigo $\bigtriangleup$),
i-PS (indigo $\diamond$),
a-PMA (red $\diamond$),
PI-75 (black box),
PBd-20 (indigo $+$),
a-PS (blue $\diamond$),
PBd-98 (black $\bigtriangleup$),
PEO (green $\ast$),
POM (black $\diamond$),
a-PHMA (black $\ast$),
a-PVA (blue $\times$),
SBR (blue $\circ$),
P6N (black $\times$),
a-P$\alpha$MS (red $\ast$),
a-PEA (green $+$),
PET (green $\diamond$),
s-PP (indigo $\ast$),
PE(413) (orange $\bigtriangleup$),
a-POA (magenta $\diamond$),
PC (red $\bigtriangleup$),
PE (orange $\times$), and
PTFE (black $\circ$).
}
\end{figure}

\begin{figure}
\includegraphics[angle=270,width=0.5\columnwidth]{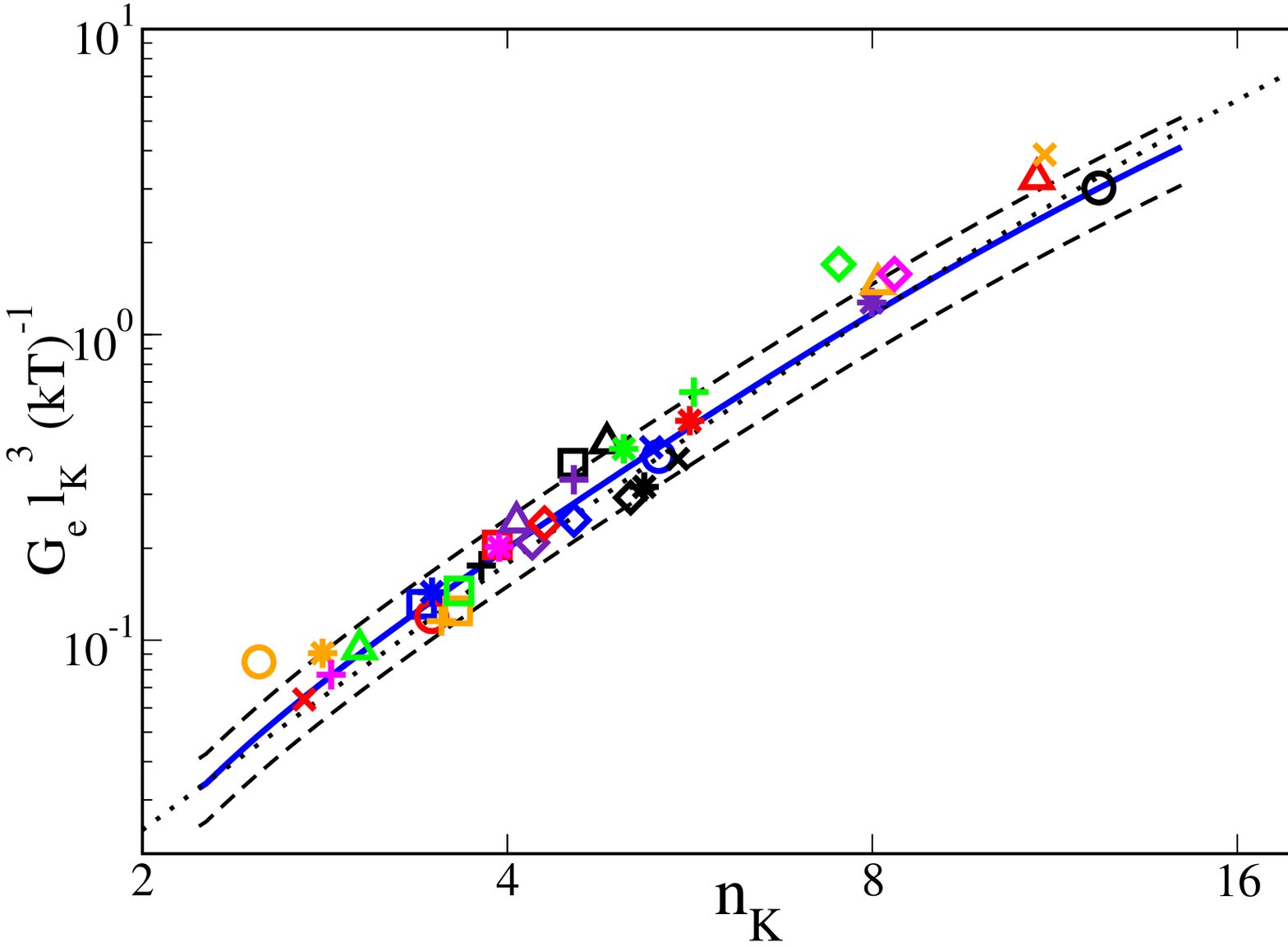}

\caption{\label{fig:reducedplateaumoduli2}Reduced entanglement moduli for the
experimental data in Fig. \protect\ref{fig:reducedplateaumoduli1} compared
to the range of KG models for $-1\leq\kappa\leq2.5$ (blue solid line),
with indications of $\pm 25\%$ error (dashed black lines). Also shown is the semi-empirical prediction
of eq. (42) in Ref. \cite{svaneborg2018KGCharacterization}) (black dotted line)
}
\end{figure}

Figure \ref{fig:reducedplateaumoduli1} shows the reduced entanglement moduli as 
a function of Kuhn number. The experimental data are in good agreement with
eq. (\ref{eq:reducedmodulus}) for flexible chains.\cite{fetters94} 
The scatter observed between the experimental plateau moduli and the predicted
plateau modulus line must be attributed either to chemical details
causing some small degree of non-universal behaviour\cite{unidad2015consequences},
such as a non-negligible crystalline fraction, or to experimental uncertainties
in accurately estimating the plateau modulus which can be quite difficult.\cite{LikhtmanMcLeishMM02}
For the very largest Kuhn numbers, the experimental data points can not discriminate between the packing argument and the predicted cross-over to the tightly entangled regime.\cite{Uchida_jcp_08,svaneborg2018KGCharacterization}

Figure \ref{fig:reducedplateaumoduli2} shows a comparison between experimental plateau moduli and entanglement moduli of KG melts extracted from Primitive Path Analysis~\cite{hou2010stress,svaneborg2018KGCharacterization}.
Most of the experimental values are within the $25\%$ error interval around the line defined by the one parameter KG models. This is concrete evidence, that the emergent entanglement properties of our KG models agree with those of the targetted experimental polymer systems.
Interestingly, the stiffer KG models also seem to be in excellent agreement with the predicted cross-over to tightly entangled regime.\cite{Uchida_jcp_08,svaneborg2018KGCharacterization}

\section{Discussion\label{sec:Discussion}}

Polymeric systems exhibit a wide range of characteristic time and length scales.
This is readily illustrated for the example of natural rubber, i.e. melts of {\it cis-}PI chains with a typical length of  $N_K=10^4$ Kuhn segments. 
Important characteristic length scales comprise (i) the Kuhn length, $l_K \approx 1nm$, (ii) the tube diameter, $d_T \approx 5nm$, (iii) the coil diameter,  $\langle R^{2}\rangle \approx 100 nm$, and (iv) the contour length, $L \approx 10 \mu m$.
The spread is even larger between the characteristic time scales.
There are already almost three orders of magnitude between the Kuhn time, $\tau_K\sim 1\times 10^{-7}s$, and the entanglement time, $\tau_e\sim 7\times 10^{-5}s$. The Rouse time of $\tau_{R}\sim 10^8 \tau_K \sim 10s$ governs {\em fast} processes such as the tension equilibration inside the tube~\cite{DoiEdwards86}, while the estimated disentanglement time is $\tau_{max}\sim 4h$.

The slow dynamics has dramatic consequences for macroscopic properties such as the viscosity. Our estimate of the effective viscosity at the Kuhn scale is $\eta_K\approx 0.4 Pa\, s$. The viscosity of a short chain melt at the entanglement threshold, $N_K=N_{eK}$, is already two orders of magnitude larger, $\eta_e\approx 40 Pa\, s$, while for our strongly entangled ($Z=N_K/N_{eK} =400$) example, $\eta \approx 5\times 10^{9} Pa\, s$. In other words, the long chain melt exhibits a {\em macroscopic} viscosity similar to glass forming liquids close to $T_g$, even though locally the chains experience a friction as if they were immersed in motor oil.

The wide range of relevant time and length scales in polymeric systems makes them natural targets for multi-scale modelling.\cite{faller2007coarse,MultiscalePeterKremerSOftMatter2009,MultiscalePeterKremerFaradayDiss2010} In particle-based models, the resolution ranges from the atom scale to DPD-like descriptions, where entire chains are represented by one or two soft spheres or ellipsoids\cite{hahn2001simulation,muller2013speeding}. What is the natural place of KG-like models in this hierarchy? And how should they be parameterized?

\begin{figure}
\includegraphics[angle=270,width=0.5\columnwidth]{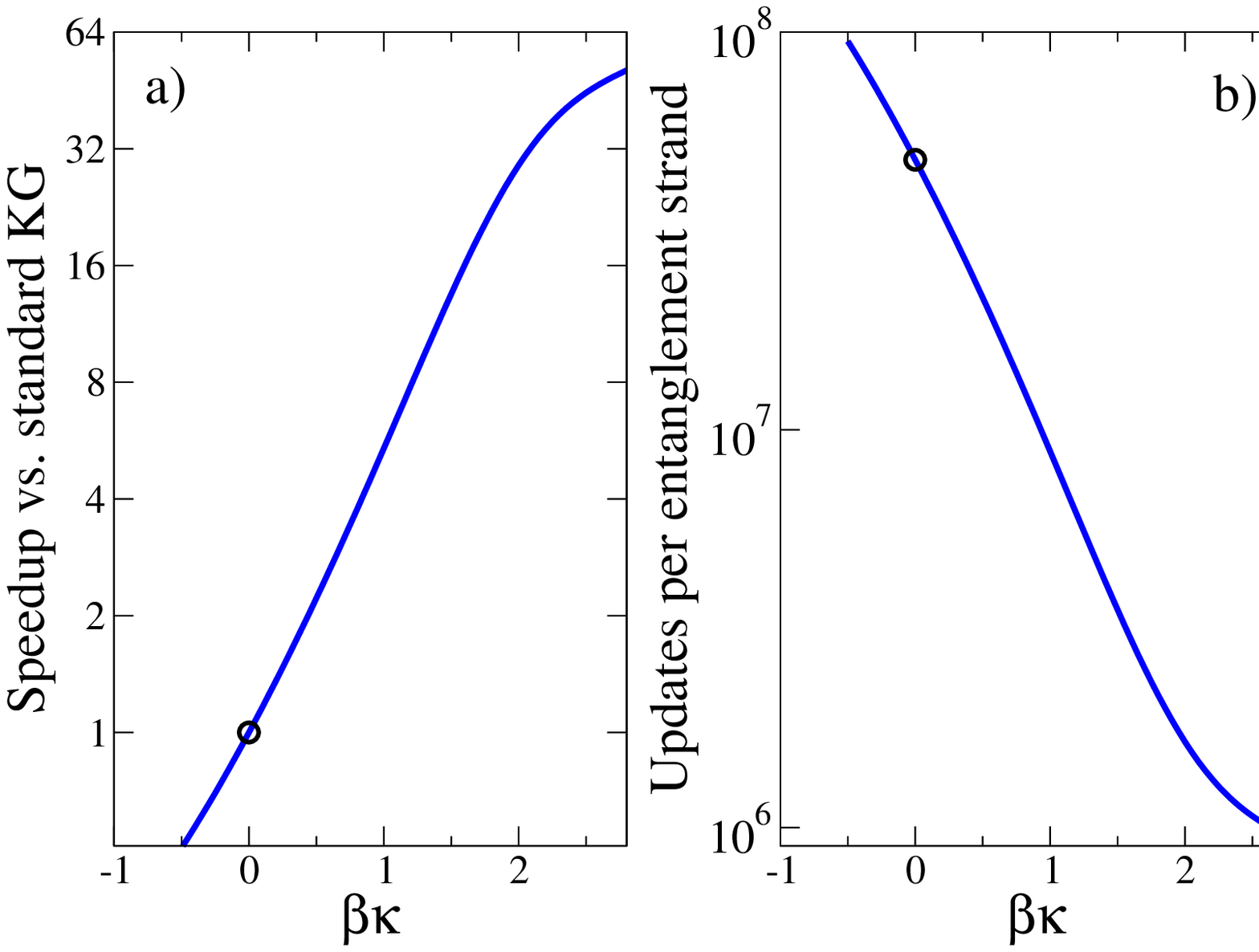}
\caption{\label{fig:speedupKG}
Speed up of KG models due to increased stiffness. The left graph shows the speed up relative to the
standard KG model with $\kappa=0$, while the right graph shows the number of particle updates, $N_{eb} \frac{\tau_e}\tau \frac\tau{\delta t}$, required to follow the dynamics of one entanglement strand over the entanglement time.
}
\end{figure}

\subsection{Entanglement vs. Kuhn scale as targets for KG models}

Typically, the KG model is mapped to experiments~\cite{kremer1990dynamics,kroeger2004simple} or simulations of more microscopic models~\cite{takahashi2017} on the entanglement scale. 
The mini-review in Sec.~\ref{sec:KuhnScale} explains in some more detail the statement from the Introduction, that one can hope to reproduce the large scale dynamics of a target system by carrying out KG simulation with chains of an appropriate effective length, $Z=L/L_e$, and then converting the results by identifying the tube diameter, $d_T$, as the unit of spatial distance as well as the entanglement time, $\tau_e$, as the unit of time. 
Conceptually, this is what universality is all about and not different from using experimental data for PDMS to predict universal aspects of the behavior of, say, amorphous polystyrene. 

%In their original publication~\cite{kremer1990dynamics}, Kremer and Grest discussed such a mapping (i.e., in the present notation, of their standard KG model with $\kappa=0$) for several polymer species. Using their estimate of the entanglement length and time as well as the tube diameter in KG units, they could estimate $\sigma$ and $\tau$ by matching experimental values.
%%of beads ($N_e=35$ extracted with a different prefactor from the $t^{1/2}$ to $t^{1/4}$ crossover of mean-square bead displacements, where eqs. (\ref{eq:lkkg}, \ref{eq:nekinterpolation}) give $N_e=75$), they could identify $\sigma$ by matching the KG tube diameter to experimental estimates from the chain statistics and plateau moduli. Finally, by matching diffusion coefficients for chains in the Rouse regime, they were able to derive a time mapping. 
%Not surprisingly, their mapping relations are similar to ours for polymer species like PI and PDMS, which have similar $n_K$ values as the standard KG model
%%{\bf $\sigma$s are comparable for all polymers PE,PS,PI where we have overlap. Also their $\tau$ for PDMS agrees with ours. Their $\tau_e$ value for PI is about x300 too small compared to our experimental value. }
%%{\bf Ralf: I don't get it. They cannot predict $\tau_e$, that's an input parameter. Need to read. Curious, I don't seem to have acces. Can you send me the .pdf?}
%% {\bf I didn't check, that's a GUESS!}. 
%%If the KG model is used to target the entanglement scale, 

Since in this view there is nothing special about the original KG model, one can apply the same logic to any other member of the family of KG models we are considering here.
As illustrated by Fig.~\ref{fig:speedupKG}, it is indeed tempting to use the additional stiffness parameter \cite{faller2000local} to reduce the CPU time required to reach the entanglement scale~\cite{faller2001chain,halverson2011molecular}. 
%Within the limits imposed by the isotropic-nematic transition [] as well as the crossover from the loose to the tight entangled regime [Uchida], considerations of efficiency would suggest the use of larger values~\cite{henderson} for the Faller and M\"uller-Plathe  [Faller & FMP] stiffness parameter to reduce the CPU time required to carry out simulations for a targeted system: 
%(i) the less beads are needed to represent an entanglement segment (see, for example, Fig. ??? in the accompanying paper~\cite{KGmodel}), the smaller the number of microscopic degrees of freedom needed to represent a system of the targeted absolute size;
%(ii) the smaller the gap between the time step of the simulation and the entanglement time (see, for example, Fig. ??? in the accompanying paper~\cite{KGmodel}), the smaller the number of time steps needed to reach the targeted simulation time.
But how far up the scales can one safely push the characteristic features of the KG model like the well-defined, almost inextensible contour length and the almost fully excluded molecular volume?
These features are adequate for a description on the Kuhn scale, but not for a generic model of loosely entangled chains at the entanglement scale.

Targeting the Kuhn scale, as we advocate here, provides a simple physical motivation for the choice of the stiffness parameter and should help to reduce ``gaps'' \cite{takahashi2017} relative to predictions of more microscopic models for the local behavior.
Importantly,  Kuhn scale-mapped KG models are in most cases computationally {\em more} efficient than the original KG model in reaching the entanglement scale, even though, by targeting the Kuhn scale, they are nominally more microscopic.The reason is that $\kappa>0$ for most Kuhn scale-mapped KG models of commodity polymers, while the original KG model maps on the intrinsically most flexible polymer species.
Typical speedups are of the order of 4, in the case of polycarbonate they reach a factor of 30 (Fig.~\ref{fig:speedupKG}).

\subsection{Linear vs. nonlinear universality in the rheology of polymer melts}

Crucially, we can hope to extend the validity range of the KG model by, to paraphrase Einstein, making the chains ``as stiff as possible, but not stiffer.'' 
By reproducing the number of Kuhn segments per entanglement length, $N_{eK}$, the models account for the maximal chain extension, $\sim\sqrt{N_{eK}}$, under strong deformations.  
Furthermore, KG melts parameterized at the Kuhn scale plausibly exhibit friction reduction in fast elongational flows, in so far as the effect can be attributed to the alignment of the Kuhn segments to the stretching direction\cite{IannirubertoBrasielloMarrucci,yaoita2012primitive}. 
There are thus good reasons to expect, that  the models discussed in the present article fulfil all {\em three} conditions for {\em non-}linear universality in the rheology of polymer melts \cite{WingstrandAlvarezHuangHassager}.

\subsection{Computational performance of Kuhn scale-mapped KG models compared to descriptions on neighboring scales}
%\subsection{Speedup of Kuhn scale-mapped KG models relative to atomistic simulations}

Kuhn scale-mapped KG models are computationally much less demanding than atomistic simulations. 
This is due to two factors:
(i) There is a considerable reduction in the number of degrees of freedom. We have not counted atoms, but assuming carbon and hydrogen atoms as the dominant components, molecular bead weights between $40 g/mol$ and $700 g/mol$ translate to $3$ to $50$ united atoms that are being represented by one KG bead. If hydrogen atoms are represented explicitly, then these numbers increase by a an additional factor of two or three. 
(ii) At the reference temperature, $T_{ref}^{dyn}$, of the rheological experiments, our estimates for the physical meaning of the KG unit of time, $\tau$, vary in the range $5ps < 1\tau < 0.35\mu s$. The corresponding time step of $50fs < \delta t = 10^{-2}\tau < 3.5ns$ is thus $50$ to $3.5\times 10^6$ times larger than the $1fs$ time step in atomistic simulations.
The time step in atomistic simulations is dictated by typical frequency of bond vibrations. Whereas if bond lengths are constrained, then the typical time scale is that of bond angle vibrations which occurs on time scales of tens of $fs$.
\cite{faller2007coarse,ShakeCiccottiFerrarioRyckaert} For systems closer to the glass transition, the speedup in modelling the large scale behaviour along the present lines would be exponentially larger. Compared to an atomistic model, this obviously comes at the price of loosing the ability to {\em predict} any of the glassy behaviour.

With at least $10^6$ particle updates per entanglement strand and time (inset Fig.~\ref{fig:speedupKG}), Kuhn scale-mapped KG models are bound to be slower than PPA-parameterized slip-link augmented DPD-models \cite{hua1998segment, masubuchi2001brownian, likhtman2005single, chappa2012translationally,TheodorouSlipSpring}. Again, the more coarse-grain description benefits from a reduction of the number of degrees of freedom by a factor of the order of $1/N_{eK}$ as well as a corresponding reduction of the number of time steps by a factor of $\tau_K/\tau_e \sim 1/N_{eK}^2$.
For the intrinsically most flexible polymers in Table~\ref{tab:Kuhn}, the speedups may be as large as a factor of $10^4$ or even $10^5$ in rare cases. 
While this approach is clearly successful, there is nevertheless a price to be paid:  effects of topological constraints do not {\em emerge} through the same mechanisms as in the target systems. These effects have to be introduced explicitly in models developed at the entanglement scale. While the tube/slip-link model is in general well understood~\cite{LikhtmanMcLeishMM02}, we suspect that non-linear universality~\cite{WingstrandAlvarezHuangHassager} or the emergence of crumpling in non-concatenated ring melts~\cite{halverson2011molecular,rosa2014ring} remain a challenge for such models.

\subsection{Kuhn scale matching as a special case of structure based coarse-graining}

The construction of coarse-grain models requires choices and the definition of (subjective) priorities. A classic example is the tension between structure-based approaches\cite{kremer2000computer,kremer2003computer,MultiscalePeterKremerSOftMatter2009} and schemes focused on preserving thermodynamic properties\cite{marrink2007martini}.

Kuhn scale matching can be viewed as a special case of structure-based coarse-graining. It is guided by theoretical considerations, which identify the Kuhn scale as controlling the emergent, universal behavior at larger time and length scales. 
Consequently, no particular effort is made to reproduce the local behavior.
The resulting ``one parameter force-field'' for the KG model is remarkably simple, but this simplicity obviously comes at the price of loosing the ability to predict (or to understand) the behaviour of experimental target systems below the Kuhn scale. In particular, this holds on the bead scale, where we employ a computationally convenient, generic model without any particular relation to the properties (or the structure) of the target system.

%In principle, the limitations of our approach can be overcome in systematic multi-scale simulations of targeted materials at particular state points  [KK reviews]. 
The techniques for structure-based coarse-graining are well understood\cite{lyubartsev1995calculation,muller2002coarse,izvekov2005multiscale,ruhle2009versatile}. If applied on a similar level of coarse-graining as our KG models (i.e. retaining a comparable number of degrees of freedom), the resulting models can be expected to offer a locally more faithful representation.  The differences are probably minor for polymers like isotactic polystyrene, where our KG beads represent three polystyrene monomers. The situation is different for polymers like polycarbonate, whose monomers are represented by several KG beads. 
In this case, the ``beads'' arising from systematic coarse-graining are neither of equal size, nor spherical or nor joined in a straight line like those of our KG models.\cite{tschop1998simulation,hahn2001simulation,muller2002coarse}
Such models may provide insight into the relation between structure, local dynamics, and the dissipation mechanisms responsible for the glassy dynamics, which is lost in our approach. In terms of computational performance, they should fall in between atomistic descriptions and Kuhn scale-mapped KG models, since they need to resolve motion on smaller time scales.

\subsection{Time scales in coarse-grain models}

There is a persistent idea in the literature \cite{kroeger2004simple} that the time scale in simulations of coarse-grain models can be inferred by standard dimensional analysis. 
The difficulty becomes clear, if we try to follow this approach on the Kuhn scale. 
The time scale $l_K \sqrt{M_K/k_BT}$ can be understood as the time required by a Kuhn segment to ballistically cover a distance comparable to its size, $l_K$, if it moves at its thermal velocity, $v_K^{th}=\sqrt{k_BT/M_K}$.
In contrast, the physically relevant Kuhn time, $\tau_K$, is controlled by the local viscosity, Eq.~(\ref{eq:etaK}), which emerges from microscopic interactions below the Kuhn scale and which is expected to display an exponential WLF temperature dependence, Eqs.~(\ref{eq:tts}) and (\ref{eq:VF Kuhn time}).

The systematic linking of times scales on different levels of spatial and temporal resolution remains a challenge. 
A conceptual framework is provided by the Mori-Zwanzig projector formalism.\cite{mori1965transport,nordholm1975systematic} Here the projection operator is defined by the choice of "slow" CG variables. The formalism provides a generalized Langevin equation (GLE) for the time evolution of the CG variables, where the effect of the "fast" variables is described by the GLE memory kernel giving rise to friction and stochastic forces applied to the slow variables.
In practice, sampling such GLE memory kernels requires simulations of the fast dynamics for fixed slow variables, which is complicated and has only been achieved relatively recently.\cite{hijon2010mori,carof2014two,jung2017iterative}

In practice \cite{fritz2011multiscale}, one often uses a mapping approach, where the time scale of the coarse-grain simulations is determined by the condition, that the coarse-grain and the microscopic model predict identical dynamics on the largest time scales accessible to the microscopic approach. 
In the present case, we have used a mapping on the Kuhn scale to estimate the physical meaning of the KG time scale $\tau$.
As shown in Table~\ref{tab:PhysicalPropertiesTime} and Figure~\ref{fig:tts}, these estimates exceed by orders of magnitude the time scale arising from the standard combination $\sigma \sqrt{M_b/[k_B T_{ref}^{dyn}]}$ of the diameter and mass $M_b$ of the KG beads with the energy scale of the model. 

This mismatch strikes us as a natural and highly desired consequence of the elimination of microscopic degrees of freedom and of the associated dissipation mechanisms. 
In principle, it is possible to preserve $\sigma \sqrt{M_b/[k_B T_{ref}^{dyn}]}$ as the definition of time by tuning the friction of a Langevin (or preferentially, DPD\cite{soddemann2003dissipative}) thermostat such that the resulting $\tau_K$ matches the experimental target value. However, this would make the simulations orders of magnitude more expensive in terms of computer time  without providing additional physical insight.

\subsection{Kuhn scale matched KG models as part of a multiscale hierarchy of polymer polymers}

In our opinion, the Kuhn scale merits to be systematically included in the hierarchy of multi-scale models of polymeric systems. Omitting it risks to mask a remarkable simplicity, which emerges from the universality of polymeric behavior.

We have focused on the KG model with bending rigidity, because it has been used in a vast number of publications as a basis for studying generic polymer and materials physics, see e.g. \cite{kremer2000computer,kremer2003computer,MultiscalePeterKremerSOftMatter2009} for reviews. 
Furthermore, there are several fast equilibration procedures~\cite{zhang2015communication,moreira2015direct,SvaneborgEquilibration2016}, which allow to build very well equilibrated, highly entangled melt configurations at relatively low computational cost.
Obviously, one could apply the same logic to bead-spring models with variable density, to models based on chains of rods rather than beads\cite{sirk2012enhanced} or to lattice models\cite{carmesin1988bond,tries1997modeling,Barkema2010,SchramBarkema2018}
as long as these capture the relevant physics of polymers.

Kuhn scale-mapped polymer models are easy to connect to neighboring scales. In the ``up''-direction, the primitive path analysis~\cite{PPA} provides a systematic link to phenomenological models describing polymers on the entanglement scale\cite{hua1998segment, masubuchi2001brownian, likhtman2005single, chappa2012translationally,TheodorouSlipSpring}.
In the ``down'' direction, they enable the generation of well equilibrated atomistic material models through fine-graining of melt configurations of a chemistry-specific KG model.\cite{carbone2010fine}
%  {\bf I suppose that Kurt has done that [citation]. But not really? Because, I think the trick is really the use of a ``chemistry-specific KG model''. Otherwise, things shouldn't work out as well in terms in molecular volume, stiffness etc.?! }
%\cite{harmandaris2006hierarchical,zhang2015communication}

%\subsection{Bottom-up vs. top-down parameterization}

The information we used here to parameterize the KG model was obtained {\em top-down} from experiment\cite{fetters2007chain}. Our aim was to provide reasonable estimates of these parameters for a wide variety of polymer species and over the entire experimentally relevant temperature range.
Alternatively, one could analyse simulations of atomistic~\cite{padding2002time,liu2013coarse,salerno2016resolving,
faller2002modeling,faller2003properties,li2011primitive,
maurel2015prediction,harmandaris2006hierarchical,chen2007viscosity,fritz2011multiscale,
karimi2008fast,eslami2011coarse,
chen2008comparison,
maurel2015prediction,
strauch2009coarse,
milano2005mapping} or mildly coarse-grain
\cite{tschop1998simulation,abrams2003combined,hess2006long}
models of target polymers at specific state points. If the purpose is solely to parameterize the present model, then it suffices to analyze the simulations along the lines of the accompanying paper~\cite{svaneborg2018KGCharacterization}. The inferred  Kuhn length, density and time are straightforward to convert into a {\em bottom-up} parameterization of the KG model, which then provides access to much larger time and length scales than the original, more microscopic model. 

\subsection{Possible applications}
Compared to the original KG model, the Kuhn scale-mapped variants are as or even more computationally efficient and can be expected to be predictive outside the linear regime. In particular, the mapping relations we provide should help to establish a direct, quantitative link to experiment.
Otherwise, the models can be profitably applied to the same broad range of complex emergent phenomena as the original KG model. For instance,  static and dynamic entanglement effects including multichain mechanisms such as
constraint release \cite{graessley82,marrucci85,RubinsteinColbyConstraintRelease,marrucci2003flow,ianniruberto2014convective} and crumpling~\cite{rosa2008structure,Halverson2012,halverson2011molecular,rosa2014ring}
as well as correlation hole effects~\cite{Wittmer_Meyer_PRL04} will naturally emerge in such models, without the description needing to be accurate on the atomic scale. 
Polydispersity, branching~\cite{grest1989relaxation}, chemical cross-linking~\cite{grest90a,duering91,svaneborg2004strain}, and network aging~\cite{svaneborg2008connectivity} are
also straightforward to include.
Furthermore, such models can be used to study effects of spatial confinement~\cite{aoyagi2001molecular} in 
thin films~\cite{pierce2009interdiffusion} and brushes~\cite{MuratGrest,murat89a,grest96a} or the addition of filler particles in composite materials~\cite{yagyu2009coarse,sussman2014entanglement}, or the welding dynamics at polymer interfaces~\cite{gersappe2002molecular,ge2014healing} to name a few examples.

\section{Conclusion\label{sec:Conclusion}}

We have argued that the Kuhn scale is a natural scale (i) to link theories, experiments and simulations of amorphous polymer melts and (ii) to target in building computational polymer models. Omitting the Kuhn scale from the hierarchy of multi-scale models risks to mask a remarkable simplicity, which emerges from the universality of polymeric behavior.

In practical terms, we have shown how to model homopolymer melts of a large variety of polymer species with an extension of the Kremer-Grest model~\cite{grest1986molecular,kremer1990dynamics}, which was originally introduced by Faller and M\"{u}ller-Plathe~\cite{faller1999local}. The force field has a single adjustable parameter, the chain stiffness. We determine this parameter by matching the (Kuhn) number of Kuhn segments per Kuhn volume, $n_K = \rho_K l_K^3$, of the target polymer species and the KG polymer model. No attempt is made to reproduce smaller scale features. Besides expressions for estimating the model parameters from experimental input, we have provided tables listing which bending stiffness to use for particular polymer species and how to translate KG into SI units. Our estimates for the mapping from simulation to physical time are based on time-temperature superposition. 

Conceptually, Kuhn scale matching can be seen as a special case of structure based coarse-graining. The choice of the structural features to be preserved is guided by theoretical considerations, which identify the Kuhn scale as controlling the emergent universal polymer behavior at larger time and length scales.

The resulting coarse-graining level is about one bead per chemical monomer or two to three beads per Kuhn segment. Kuhn scale-mapped KG models thus fall in between atomistic or mildly coarse-grain models and descriptions on the entanglement scale. Both coarse-graining steps, from the atom to the Kuhn and from the Kuhn to the entanglement scale, are associated with performance gains of several orders of magnitude. For systems close to the glass transition, the speedup in modelling the large scale behaviour is even exponentially larger. 
Compared to atomistic descriptions, Kuhn scale-mapped KG models loose the ability to predict the microscopic (glassy) dynamics or to reproduce semi-crystalline ordering. 
Compared to phenomenological entanglement models, Kuhn scale-mapped KG models preserve the emergence of the full spectrum of universal amorphous polymer properties through the same mechanisms as in the experimental target systems.
In particular, we expect them to automatically fulfil all {\em three} conditions for non-linear universality in the rheology of polymer melts~\cite{WingstrandAlvarezHuangHassager}.

An interesting challenge for future work would be the parameterization of a corresponding force-field for co-polymer systems, using for example the technique from Ref.~ \cite{MilnerChiParameterPRL}. While this should, in principle, be possible at least for static properties, modelling the dynamics might no longer be as simple as adjusting a single time scale. Similarly, it might be possible to parameterise minimal models of glassy\cite{bennemann1999molecular,baschnagel2005computer} or semi-crystalline\cite{meyer2001formation} polymers along the present lines. 

\appendix
\section{Time-temperature superposition}\label{sec:TTS}

For a TTS reference temperature $T_0$ the empirical Williams-Landel-Ferry (WLF)\cite{WilliamsLandelFerry} shift factor  has the form
\begin{equation}\label{eq:WLF}
\log_{10} a_T(T;T_0)=-\frac{C_1(T_0)(T-T_0)}{C_2(T_0)+(T-T_0)}\ .
\end{equation}
Eq.~(\ref{eq:WLF}) is valid above the glass transition temperature in the temperature range $[T_g, T_g+100K]$.
Using $T_g$ as reference temperature, the constants adopt ``universal'' values $C_1^g \approx 15$ and $C_2^g \approx 50 K$.\cite{WilliamsLandelFerry} Other choices require suitably adjusted parameters $C_1(T_0)$ and $C_2(T_0)$. 

The conversion can be avoided by writing the shift factor in a form derived from the equivalent Vogel-Fulcher-Tammann-Hesse equation~\cite{vogel1921,fulcher1925,tammanhesse1926}.
\begin{equation}\label{eq:VF}
\ln a_T(T;T_0)=-\frac{C_{VF}(T-T_0)}{(T_0-T_{VF})(T-T_{VF})}\ .
\end{equation}
The relations
\begin{eqnarray}
\label{eq:TVF_WLF_conversion}
\lefteqn{T_{VF} = T_0-C_2(T_0)} \\
 &&\Leftrightarrow C_2(T_0) = T_0- T_{VF} \nonumber\\
\lefteqn{C_{VF} = \ln(10) C_1(T_0) C_2(T_0)}\\
 &&\Leftrightarrow C_1(T_0) = \frac{C_{VF}}{\ln(10) (T_0- T_{VF})} \nonumber
\end{eqnarray}
with $C_{VF} \approx 2000 K$ allow to pass between the two representations. 
In particular, Eqs.~(\ref{eq:WLF}) and (\ref{eq:VF}) suggests (i) that the viscosity diverges at the Vogel-Fulcher temperature $T_{VF}=T_g-C_2^g$ located $\sim50K$ below the glass transition temperature and (ii)  that the orders of magnitude by which the viscosity drops in the opposite limit of $T\rightarrow\infty$ are given by $C_1(T_0)$ and are hence inversely proportional to the distance of the reference from the Vogel-Fulcher temperature. 
%is  $C_1^g\approx15$ orders of magnitude smaller than at $T_g$.
More detailed information and specific tables can be found in Ref.~\cite{ngai2007temperature}. Note, however, that their Eq. (26.3) ought to read
\begin{equation}\label{eq:correction Ngai}
\log_{10} a_T(T;T_0)=\frac{C_{VF}/\ln(10)} {T-T_{VF}} - \frac{C_{VF}/\ln(10)}{T_0-T_{VF}} \ .
\end{equation}

\begin{acknowledgement}
Acknowledgement: The  simulations  in  this  paper  were  carried  out  using  the
LAMMPS molecular dynamics software.\cite{plimpton1995fast}
Computation/simulation for the work described in this paper was supported
by the DeiC National HPC Center, University of Southern Denmark, Denmark. 
\end{acknowledgement}

\providecommand{\latin}[1]{#1}
\makeatletter
\providecommand{\doi}
  {\begingroup\let\do\@makeother\dospecials
  \catcode`\{=1 \catcode`\}=2 \doi@aux}
\providecommand{\doi@aux}[1]{\endgroup\texttt{#1}}
\makeatother
\providecommand*\mcitethebibliography{\thebibliography}
\csname @ifundefined\endcsname{endmcitethebibliography}
  {\let\endmcitethebibliography\endthebibliography}{}

\end{document}